\documentclass[twocolumn]{aastex63}

\graphicspath{{./}{figures/}}

\submitjournal{ApJ}

\shorttitle{1453}
\shortauthors{Liepold et al.}

\newcommand{\ngc}{NGC~1453}
\newcommand{\mdm}{\ensuremath{M_{15}}}
\newcommand{\kms}{\ensuremath{}{\rm \, km~s^{-1}}}
\newcommand{\msun}{\ensuremath{M_{\odot}}}
\newcommand{\mbh}{\ensuremath{M_\mathrm{BH}}}

\newcommand{\ml}{\ensuremath{M^*/L_{\rm F110W}}}
\newcommand{\vc}{\ensuremath{V_c}\ }
\newcommand{\matthewpaper}{Quenneville et al., in prep }

\let\tablenum\relax
\usepackage{siunitx}
\usepackage{amsmath}

\begin{document}

\title{The MASSIVE Survey - XV. A Stellar Dynamical Mass Measurement of the Supermassive Black Hole in Massive Elliptical Galaxy NGC~1453}

\correspondingauthor{Chris Liepold}
\email{cmliepold@berkeley.edu}

\author{Christopher M. Liepold}
\affiliation{Department of Astronomy, University of California, Berkeley, CA 94720, USA}
\affiliation{Department of Physics, University of California, Berkeley, CA 94720, USA}

\author{Matthew E. Quenneville}
\affiliation{Department of Astronomy, University of California, Berkeley, CA 94720, USA}
\affiliation{Department of Physics, University of California, Berkeley, CA 94720, USA}

\author{Chung-Pei Ma}
\affiliation{Department of Astronomy, University of California, Berkeley, CA 94720, USA}
\affiliation{Department of Physics, University of California, Berkeley, CA 94720, USA}

\author{Jonelle L. Walsh}
\affiliation{George P. and Cynthia Woods Mitchell Institute for Fundamental Physics and Astronomy, and Department of Physics and Astronomy, \\
Texas A\&M University, College Station, TX 77843, USA}

\author{Nicholas J. McConnell}
\affiliation{Institute for Scientist and Engineer Educators, University of California, Santa Cruz, CA, USA}

\author{Jenny E. Greene}
\affiliation{Department of Astrophysical Sciences, Princeton University, Princeton, NJ 08544, USA}

\author{John P. Blakeslee}
\affiliation{Gemini Observatory, Casilla 603, La Serena, Chile}

\begin{abstract}
We present a new stellar dynamical mass measurement (\mbh)\ of the supermassive black hole (SMBH) in \ngc, a fast-rotating massive elliptical galaxy in the MASSIVE survey.
We measure stellar kinematics in 135 spatial bins in the central 1.5 kpc by 2 kpc region of the galaxy using high signal-to-noise ($S/N \sim 130$) spectra from the Gemini-North GMOS integral-field spectrograph (IFS). Combining with wide-field IFS kinematics out to $\sim 3$ effective radii and stellar light distributions from \emph{Hubble Space Telescope} (\emph{HST}) WFC3 images, we perform Schwarzschild orbit-based mass modeling in the axisymmetric limit to constrain the mass components in NGC~1453.
The best-fit black hole mass is $\mbh =(2.9 \pm 0.4)\times 10^9 M_\odot$; the mass models without a central black hole are excluded at the $8.7\sigma$ level.
The \ngc\ black hole lies within the intrinsic scatter of the SMBH and galaxy scaling relations, unlike three other galaxies hosting $\ga 10^{10} M_\odot$ SMBHs in the MASSIVE sample.
The high-$S/N$ GMOS spectra enable us to determine 8 moments of the Gauss-Hermite expansion of the line-of-sight velocity distributions (LOSVDs), which are used as constraints in the orbit modeling.  The stellar orbits in the mass models are further constrained to produce negligible $h_9$ through $h_{12}$ to minimize spurious behavior in the LOSVDs.
We show that truncating the series at $h_4$, as was often done in prior work, leads to a much weaker constraint on the inferred \mbh\ for \ngc.
Furthermore, we discuss precautions and modifications that are needed to achieve axisymmetry in triaxial orbit codes that use the Schwarzschild method to sample the start space of stellar orbits in triaxial gravitational potentials. 

\end{abstract}


\section{Introduction}

Making a direct dynamical measurement of the mass of a supermassive black hole (SMBH) using stellar or gas kinematics requires both exquisite observational datasets and extensive theoretical modeling.  Over three decades of efforts by multiple research groups have accumulated about 100 dynamically determined masses for SMBHs at the centers of local galaxies out to a distance of about 120 Mpc, with varying degrees of accuracy (see compilations in, e.g., \citealt{McConnellMa2013, KormendyHo2013, Sagliaetal2016}).

The high-mass regime ($\mbh \ga 10^{9.5} M_\odot$) faces the additional challenge that the host galaxies are massive elliptical galaxies whose central stellar light profiles typically have flattened cores that differ significantly from the high-density and cuspy centers of less massive elliptical galaxies and bulges of disk galaxies. 
These stellar cores are a defining feature of the most massive ellipticals (e.g., \citealt{Faberetal1997,
Laueretal2007,Coteetal2007,ferrarese2006,Grahametal2003}), indicating a significant deficit of stars, possibly due to three-body gravitational slingshots that scatter stars passing close to a SMBH binary to larger radii (e.g., \citealt{Begelmanetal1980}).
These diffuse cores make it extremely difficult to obtain stellar absorption-line spectra of high-$S/N$ quality that is needed for reliable \mbh\ measurements.  Long-integration observations on large ground-based telescopes in excellent seeing conditions or with the assistance of adaptive optics are required.

In pursuit of a comprehensive study of the highest-mass regime of local SMBHs and galaxies, we have been conducting a volume-limited survey, MASSIVE, of the most massive galaxies in the local universe \citep{Maetal2014}.
The MASSIVE survey targets $\sim 100$ early-type galaxies (ETGs) in the northern sky ($\delta > -6$ degrees) within a distance of 108 Mpc. 
Within this volume, it is designed to be complete to an absolute $K$-band magnitude of $M_K = -25.3$ mag, covering all ETGs with stellar mass $M^* \gtrsim 10^{11.5} \msun$ and with no selection cuts on galaxy size, velocity dispersion or environment. This parameter range is unexplored by ATLAS$^{\rm 3D}$, the previous volume-limited survey of 260 local ETGs out to a distance of 42 Mpc \citep{cappellarietal2011}.

We have obtained comprehensive spectroscopic data using IFS on both sub-arcsecond and arc-minute scales
and performed uniform measurements of the spatially-resolved kinematics.
Many results on the stellar kinematics and stellar populations of MASSIVE galaxies out to a few effective radii from our wide-field IFS observations can be found in \citet{Vealeetal2017a, Vealeetal2017b, Vealeetal2018, Eneetal2018, Greeneetal2015, Greeneetal2019}. 
Results from finely-resolved stellar kinematics in the central $\sim 2$ kpc regions of 20 MASSIVE galaxies are presented in \citet{Eneetal2019, Eneetal2020}. 
In addition to the IFS data, we have also assembled an extensive array of multi-wavelength data of MASSIVE galaxies to study stellar light profiles \citep{Goullaudetal2018},
cold molecular gas \citep{Davisetal2016, Davisetal2019}, warm ionized gas \citep{Pandyaetal2017}, and hot X-ray gas \citep{Gouldingetal2016, Voitetal2018}.

In addition to studying the luminous baryonic components in massive ETGs, one major science goal of the MASSIVE survey is to perform 
simultaneous dynamical mass modeling of the SMBH, stars, and dark matter for a sample of cleanly selected massive ETGs using a uniform set of sub-arcsecond and wide-field IFS data and photometric data.
To date, only 7 of the 100 galaxies in the MASSIVE survey have published SMBH masses that are determined from orbit mass modeling of stellar kinematic data. 
Three of the 7 galaxies are in the Virgo cluster: NGC~4486 (M87; \citealt{Gebhardtetal2011}; see also \citealt{walshetal2013} and \citealt{EHT2019a}), NGC~4472 (M49; \citealt{Ruslietal2013}), and NGC~4649 (M60; \citealt{ShenGebhardt2010}).  
Two others are the brightest cluster galaxies of rich clusters:  NGC~4889 in the Coma cluster and NGC~3842 in the Leo cluster \citep{McConnelletal2011b, McConnelletal2012}.
The remaining two are the brightest galaxies in galaxy groups: NGC~1600 in a fossil-like group \citep{Thomasetal2016} and NGC~7619 in the Pegasus group \citep{Ruslietal2013}.
Except for NGC~4649, the spectroscopic observations were all conducted with IFS on 8-10 meter telescopes. 
The measured \mbh\ spans an order of magnitude from $\sim 2\times 10^9 M_\odot$ to $\sim 2\times 10^{10} M_\odot$. 
More \mbh\ measurements in this mass range are clearly needed to quantify more robustly the upper end of the \mbh-galaxy scaling relations for a better understanding of black hole feedback processes and massive galaxy evolution.
We have acquired the spectroscopic and photometric data that are needed to perform dynamical modeling for the 20 galaxies reported in \citet{Eneetal2019} and several other galaxies in the MASSIVE survey.

We turn to this goal in this paper and report the stellar dynamical measurement of the mass of a new SMBH at the center of the massive elliptical galaxy \ngc, a fast rotator in the MASSIVE survey.
\ngc\ is the brightest galaxy in its galaxy group, a typical environment for MASSIVE galaxies \citep{Vealeetal2017b}.
As listed in Table~3 of \citet{Maetal2014},
the 2MASS ``high-density contrast''group catalog \citep{crooketal2007} identified 12 galaxies as members in the \ngc\ group, and estimated the virial mass of the group to be $10^{13.9} M_\odot$, presumably with large errors due to the small number of member galaxies.
Our \emph{HST} images of \ngc\ show very regular elliptical isophotes (Figure~13 of \citealt{Goullaudetal2018}).
The photometric and kinematic axes are also closely aligned \citep{Eneetal2018, Eneetal2019, Eneetal2020}, suggesting that the galaxy can be approximated as an axisymmetric system.

A distance measurement is needed to convert the observed angular scales to physical length and mass scales, and the inferred \mbh\ scales linearly with the assumed distance.
For \ngc, we use our new determination of 51.0 Mpc from the MASSIVE-WFC3 project \citep{Goullaudetal2018} using the surface-brightness fluctuation technique (Jensen et al. in prep).  This new distance is about 10\% smaller than 56.4 Mpc from group-corrected flow velocity in the 2MASS redshift survey. For a flat $\Lambda$CDM with a matter density of $\Omega_m = 0.315$ and a Hubble parameter of $H_0=70 \kms$ Mpc$^{-1}$, 1 arcsec is 245 pc at 51.0 Mpc.

We perform Schwarzschild orbit modelling \citep{Schwarzschild1979} using the triaxial implementation described by \citet{vandenBoschetal2008}. We perform this modelling in the axisymmetric limit, and in Section~\ref{axisymmetric_limit} provide a prescription for how to achieve this limit properly in the triaxial code.  The line-of-sight stellar velocity distributions (LOSVDs) are the main observational inputs in any stellar dynamical mass modeling of galaxies using orbit-based methods. It is a common practice, and the practice within this code, to expand the LOSVDs in a Gauss-Hermite series \citep{vanderMarelFranx1993,Rixetal1997}.
The Gauss-Hermite expansion provides a natural way to express deviations from a Gaussian distribution since the terms in the series are orthogonal and linear.  However, there has been little discussion in the literature about the appropriate order at which to truncate the series.  To date, most published work on \mbh\ measurements that relied on the Gauss-Hermite expansion of the LOSVDs had measured only the lowest four moments from the stellar spectra (i.e., velocity $V$, dispersion $\sigma$, skewness $h_3$, and kurtosis $h_4$), using only these moments as observational constraints in subsequent orbit modeling and ignoring all higher moments.  In this paper, we investigate the importance of including the higher moments for constraining \mbh\ in \ngc. When higher moments are left unconstrained, the LOSVDs predicted by the orbit models can contain large spurious contributions from these high moments.
 
In Sec.~2, we describe the spectroscopic observations and the resulting stellar kinematics from the Gemini Multi-Object Spectrograph (GMOS; \citealt{Hooketal2004}) IFS of the central $\sim$1.5 kpc by 2 kpc region of \ngc\, and the wide-field coverage with the McDonald Mitchell IFS \citep{hilletal2008a}.  
In Sec.~3, we describe our IR imaging observations of \ngc\ from the \emph{HST} Wide Field Camera 3 (WFC3) and
the determination of the 2D light profile and the 3D deprojected stellar mass profile.
The orbit modeling method is discussed in Sec.~4.
The mass modeling results are given in Sec.~5,
and the best-fit mass model is discussed further in Sec.~6.
In Sec.~7, we discuss a number of relevant issues: the impact of Gauss-Hermite series truncation on the inferred \mbh, the subtleties in achieving axisymmetry within the triaxial code, comparisons to results from Jeans modelling, implications for the black hole scaling relations, and connections to our previous observations of warm ionized gas in \ngc\ \citep{Pandyaetal2017}.

\section{Spectroscopic Data and Stellar Kinematics}

As part of the MASSIVE survey, 
we obtained spatially-resolved stellar spectra for NGC~1453 with the Gemini Multi Object Spectrograph (GMOS, \citealt{Hooketal2004}) in the IFS mode on the 8.1 m Gemini North Telescope
and the Mitchell/VIRUS-P IFS \citep{hilletal2008a} on the 2.7 m Harlan J. Smith Telescope at McDonald Observatory.  
Here we summarize the observations, data reduction processes, and the procedures used to extract the stellar kinematics.

\subsection{Central kpc kinematics} 
\label{central_kpc_kinematics}

We observed the central $\sim$1.5 kpc $\times\,2$ kpc region of \ngc\ 
using GMOS in the 2015B semester.
The two-slit mode of GMOS provided a field of view of  $5'' \times 7''$ consisting of 1000 hexagonal lenslets, each with a projected diameter of $0.2''$.  
An additional 500 lenslets observed simultaneously a $5'' \times 3.5''$ region of the sky, which was offset by about $1'$ from the science field. 
The R400-G5305 grating and CaT filter combination was used to avoid spectral overlap on the detector and to provide a clean wavelength coverage of 7800-9330 \AA. 
The spectral resolution of GMOS is determined from arc lamp lines for each lenslet with a mean value is 2.5 \AA\ FWHM.
Six science exposures, each of 850 seconds, were taken.  
The median seeing was $0.7''$ FWHM.
Other details and our data reduction procedure are described in \citet{Eneetal2019}.

\begin{figure}
  \includegraphics[width=\columnwidth]{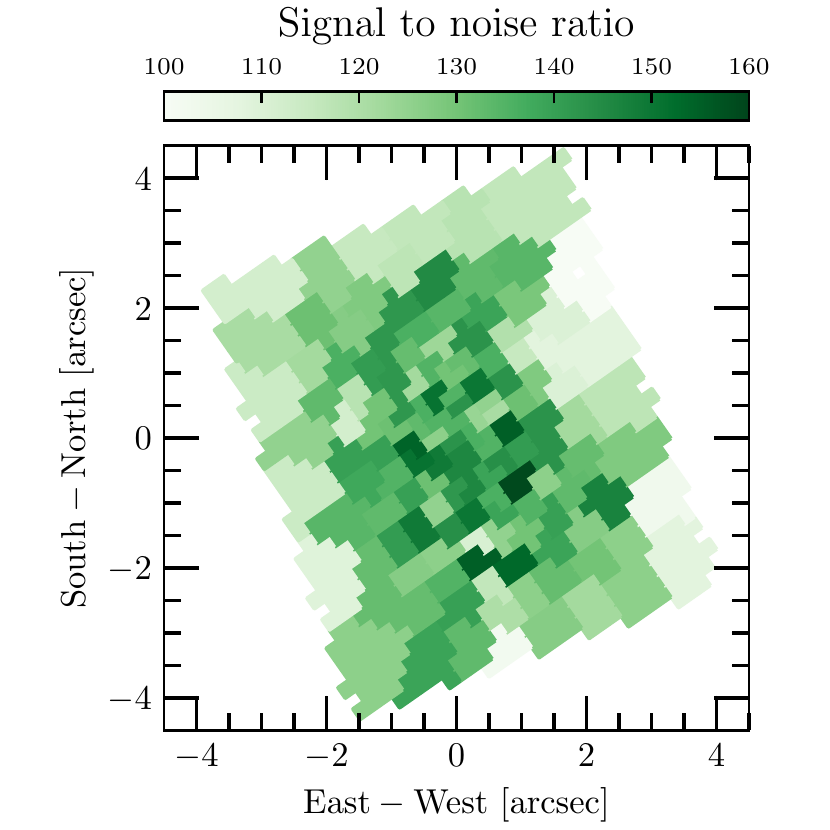}
    \caption{ Signal-to-noise map of the Gemini GMOS IFS data for the 135 Voronoi bins in the central $5'' \times 7''$ of NGC~1453.  A target $S/N$ of 125 is used in the binning procedure.
    The $S/N$ value for each bin scatters around the target with a typical RMS scatter of $\sim 10\%$, while the innermost bins achieve $S/N$ up to $\sim 150$.
       Stellar kinematics from high-quality spectra are critical for measuring the gravitational effects of the central black hole. 
     Our observations are able to achieve this high $S/N$ over finely resolved spatially bins; both needed for kinematic extraction and black hole measurements.
    }
    \label{fig:N1453_binmap}
\end{figure}
\begin{figure}
  \includegraphics[width=\columnwidth]{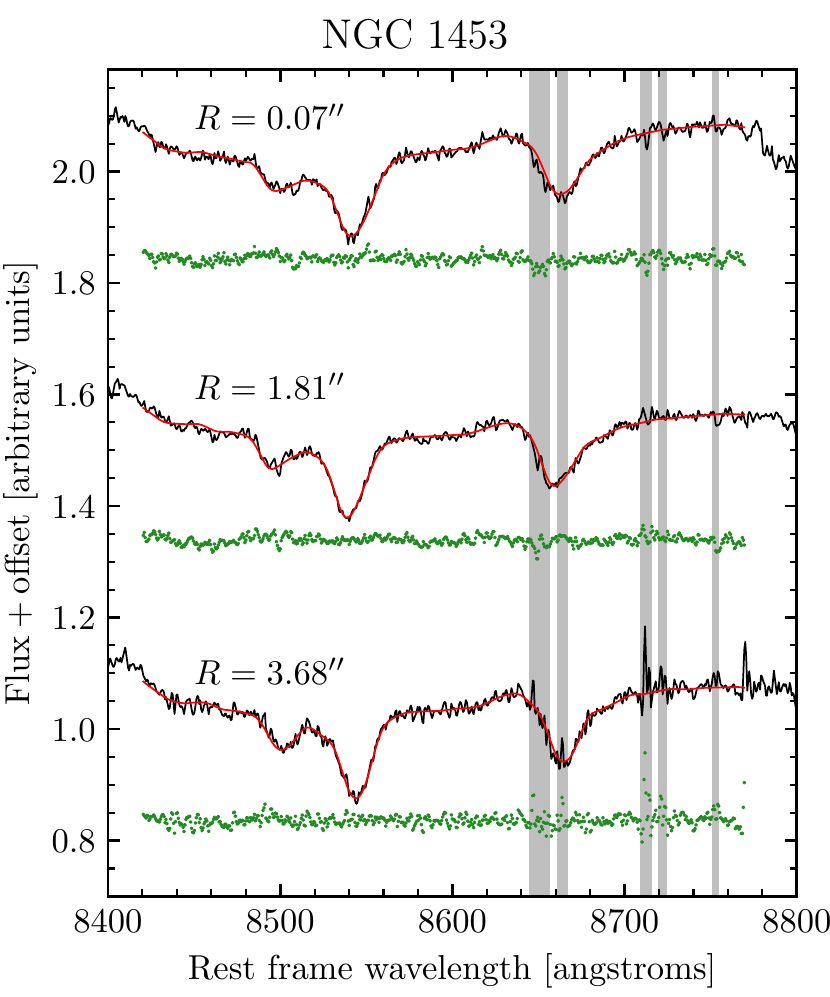}
    \caption{
    CaII triplet region of the Gemini GMOS IFS spectra (black) for three representative bins at three locations of NGC~1453: center with $S/N = 143$ (top), $1.81''$ from center with $S/N = 130$ (middle), and $3.68''$ from center with $S/N = 112$ (bottom).  The stellar template broadened by the best-fit LOSVD (red) is overlaid on each observed spectrum.   The fit is performed over the rest wavelength range of 8420 -8770 \AA\ centered around the CaII triplet absorption lines, excluding the grey shaded regions of improperly subtracted sky lines. The fit residuals (green dots) are shifted by an arbitrary amount for clarity. 
    }
    \label{fig:N1453_gmos}
\end{figure}

We use the CaII triplet absorption features over the rest wavelength range of 8420-8770 \AA\ to measure the stellar kinematics.
We apply the Voronoi binning algorithm \citep{CappellariCopin2003} with a target $S/N$ of 125 to determine how to spatially group the individual GMOS lenslets to achieve uniformly high-quality spectra.  
The procedure returns $S/N$ values (per spectral pixel of 0.67 \AA) that scatter about the target with an rms of $\sim 10$\%. 
Spectra from individual lenslets within a Voronoi bin are co-added as described in \citet{Eneetal2019}. 
After fitting the spectra with pPXF, we re-estimate the $S/N$ as the ratio of the median flux and the root-mean-square residual from the fit.
The resulting $S/N$ map for the 135 Voronoi bins is shown in Figure~\ref{fig:N1453_binmap}.
The resulting CaII region of the spectra for three representative bins are shown (black curves) in Figure~\ref{fig:N1453_gmos}.

\begin{figure*}
  \includegraphics[width=7in]{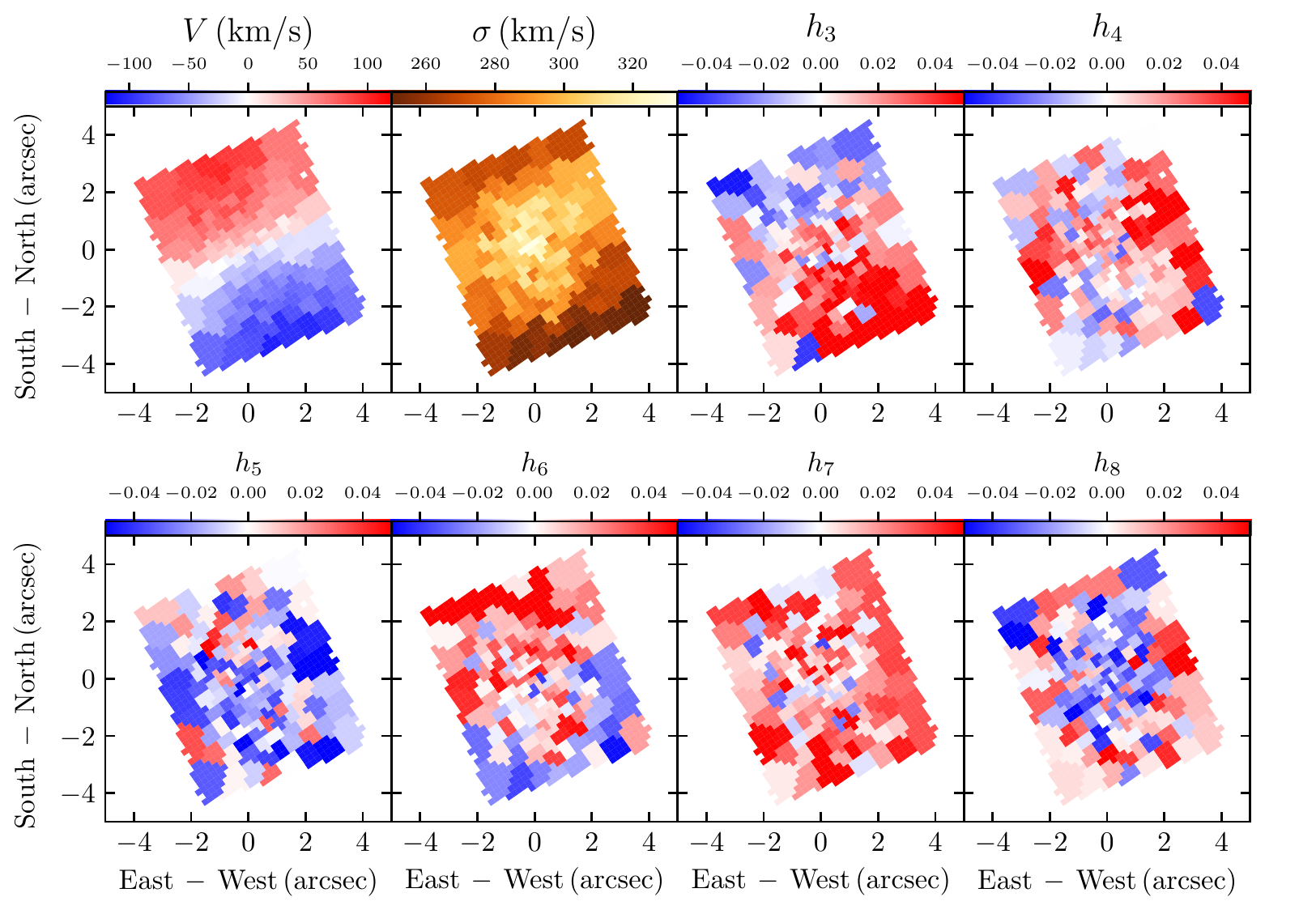}
    \caption{
    Maps of the stellar kinematics measured from the Gemini GMOS IFS over 135 spatial bins in the central $5''\times 7''$ of \ngc.  Each panel shows one of the eight velocity moments in the Gauss-Hermite expansion of the LOSVD: velocity $V$, velocity dispersion $\sigma$, and the $h_3$ to $h_8$ higher moments.  The velocity map shows a regular rotation pattern with $|V|$ reaching $\sim 100\kms$, and the $\sigma$ map shows a central peak. North is up and east is to the left.
    }
\label{fig:gmos_map}
\end{figure*}

We measure the stellar line-of-sight velocity distribution (LOSVD) within each spatial bin
using the penalized pixel-fitting (pPXF) method of \citet{Cappellari2016}.
The LOSVD is parameterized as a Gauss-Hermite series\footnote{Note that the pPXF method described in \citet{CappellariEmsellem2004} only allows $n = 2$, $4$, or $6$. The version described in \citet{Cappellari2016} allows arbitrary $n$.} up to order $n$
\begin{equation}
\label{eq:GH}
f(v) = \frac{e^{-\frac{y^2}{2}}}{\sqrt{2 \pi \sigma^2}} \bigg[1 + \sum_{m=3}^n h_m H_m(y) \bigg],
\end{equation}
 where $y=(v-V)/\sigma$, $V$ is the mean velocity, $\sigma$ is the velocity dispersion, and $H_m$ is the $m^{\text{th}}$ Hermite polynomial as defined in Appendix A of \cite{vanderMarelFranx1993}.
 
For each spectrum, the stellar continuum is modeled with an additive polynomial of degree zero (i.e., an additive constant) and a multiplicative polynomial of degree three.
A set of stellar template spectra are convolved with the instrumental line spread function and the LOSVD before adding and multiplying by these polynomials.
The polynomial coefficients, template weights,
and Gauss-Hermite moments are fitted simultaneously.  

To test for potential issues with template mismatches, we compare two sets of stellar templates 
chosen from the Calcium Triplet (CaT) Library of 706 stars \citep{Cenarroetal2001} and find negligible differences in the resulting kinematics.
The first set contained 15 stellar templates of the same 15 stars used in the extensive tests in \citet{Barthetal2002}.
For the second set, we use all 360 G and K stars in the CaT Library for each bin. 
The resulting $V$ and $\sigma$ differ by an average of $\sim 5 \kms$ and the higher moments by $\sim 0.01$, all well within the measurement errors.
Our kinematic moments determined from the CaII triplet region are therefore robust to template choices, similar to the findings in \citet{Barthetal2002}.
The stellar spectra of the CaT library cover the wavelength range of 8348-9020 \AA\ with a spectral resolution of 1.5 \AA\ FWHM. 

The resulting stellar template broadened by the best-fit LOSVD is shown for each of the three example bins in Figure~\ref{fig:N1453_gmos} (red curves).  
We use a bootstrap approach to determine the error bars on the kinematic moments of each LOSVD.  
For comparison, we have also estimated the errors using the standard Monte Carlo method with 100 trial spectra per bin.
The bootstrapped errors on the kinematic moments are typically 50\% to 100\% larger than the Monte Carlo errors. 
See Sec. 4 of \citet{Eneetal2019} for a detailed discussion.

The maps of the 8 kinematic moments, $V$, $\sigma$, $h_3$, $\dots$, $h_8$, are shown in Figure~\ref{fig:gmos_map}. 
The velocity map shows a regular rotation pattern with $|V|$ reaching $\sim 100\kms$, and the $\sigma$ map shows a central peak of $\sim 325\kms$.
The mean errors are $\SI{7.1}{\kilo\meter\per\second}$ for $V$ and $\SI{8.4}{\kilo\meter\per\second}$ for $\sigma$. The mean errors for $h_3$ through $h_8$ are quite similar, varying from 0.018 to 0.023.
The radial profiles of these moments are shown below in Figure~\ref{figure:kinematics_radial}.

\begin{figure*}[ht]
  \centering
  \includegraphics[width=0.9\linewidth]{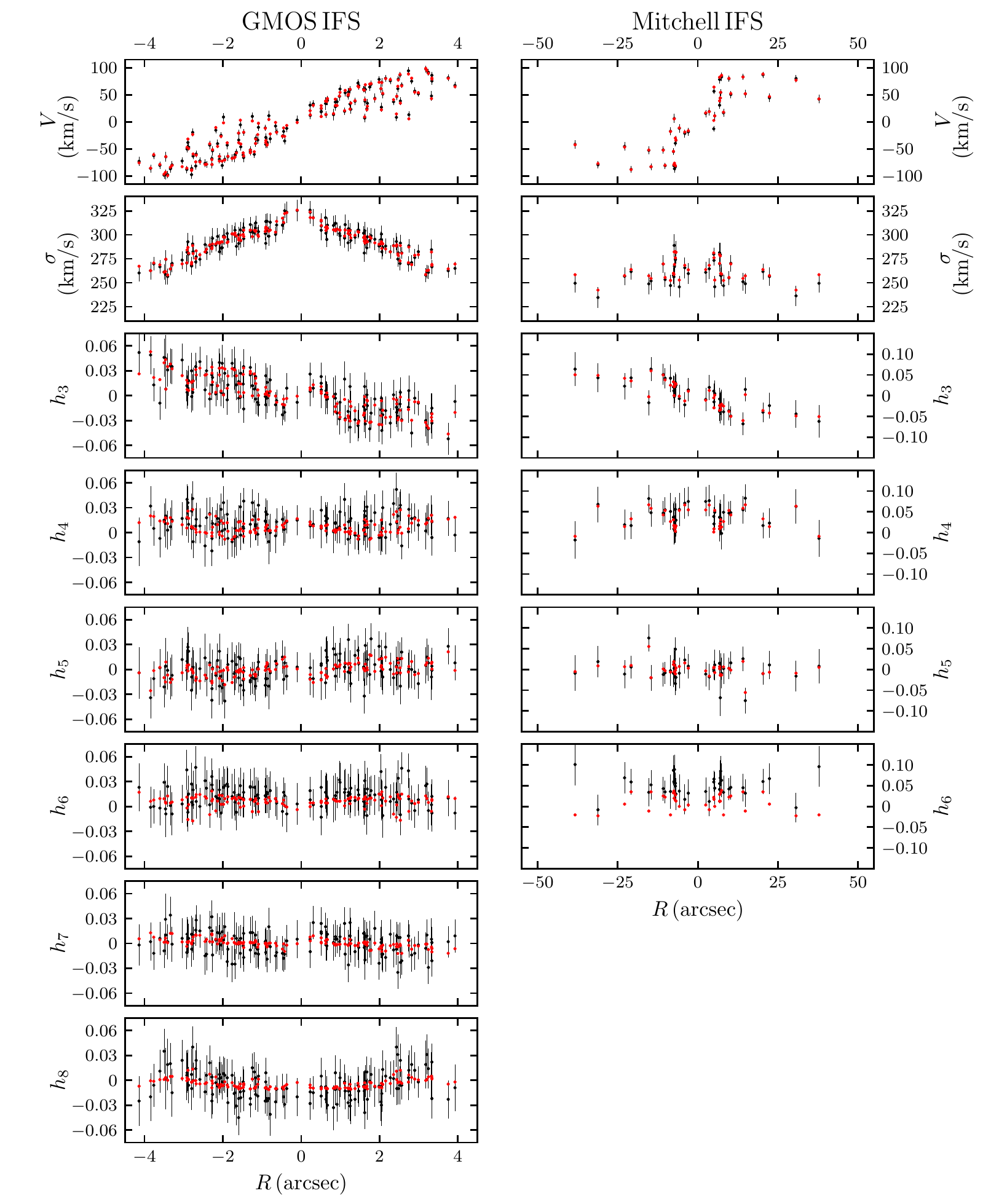}
  \hfill
  \caption{(Unfolded) radial profile of the stellar kinematics
determined from GMOS (left) and Mitchell (right) observations (black), and kinematics predicted by the best-fit mass model (red) with black hole mass $\mbh = \SI{2.9e9}{}M_\odot$, stellar mass-to-light ratio $\ml = 2.09$ (in solar units), and enclosed dark matter halo mass (within 15 kpc) $\mdm = 7 \times 10^{11} M_\odot$. 
The kinematic bins have been unfolded so that bins whose centers lie between $-90^\circ$ and $+90^\circ$ of the photometry PA are plotted with positive $R$ and others are shown with negative $R$. 
The rotation in $V$ and central values of $\sigma$ are well-fit by this model, and the high moments $h_5 - h_8$ are close to 0 with some scatter.  
\label{figure:kinematics_radial}}
\end{figure*}

\subsection{Wide-field kinematics}
\label{wide_field_kinematics}

We observed NGC~1453 as one of the 100 MASSIVE galaxies in 2013 trimester 3, using the Mitchell/VIRUS-P IFS.
The Mitchell IFS consists of 246 evenly spaced fibers with a one-third filling factor. Each fiber has a $4.1''$ diameter, and the IFS covers a large $107''\times 107''$ FOV.
Three dither positions of equal exposure time were used to obtain contiguous coverage of NGC~1453.  
We interleaved a 10-minute exposure on sky and two 20-minute exposures on target for a 2-hour total on-source exposure time.  
The spectral range spans 3650-5850 \AA, covering the Ca HK region, the G-band region, H$\beta$, Mgb, and several Fe absorption features.
 
Individual central fibers have $S/N$ above 50, while the outer fibers are binned spatially to achieve a $S/N$ threshold of 20 for the fainter outskirt of the galaxy. 
A similar procedure as in Sec.~2.1 is used to determine the stellar LOSVD for each of the 38 spatial bins.
We used the MILES library of 985 stellar spectra (\citealt{Sanchez-Blazquezetal2006, Falcon-Barrosoetal2011}) as stellar templates and ran pPXF over the full library for each spectrum. 
Further details are described in \citet{Maetal2014} and \citet{Vealeetal2017b}.

As can be seen in Figure~\ref{figure:kinematics_radial} here and Figure~21 of \citet{Eneetal2019}, the kinematic moments in the innermost Mitchell bins match well with the GMOS moments.

\section{Photometric Data} \label{photometric_data}

\begin{figure*}
  \centering

	\begin{tabular}{cc}
	\begin{minipage}[t]{0.3293\textwidth}
	\vspace{0pt}\includegraphics[width=\textwidth]{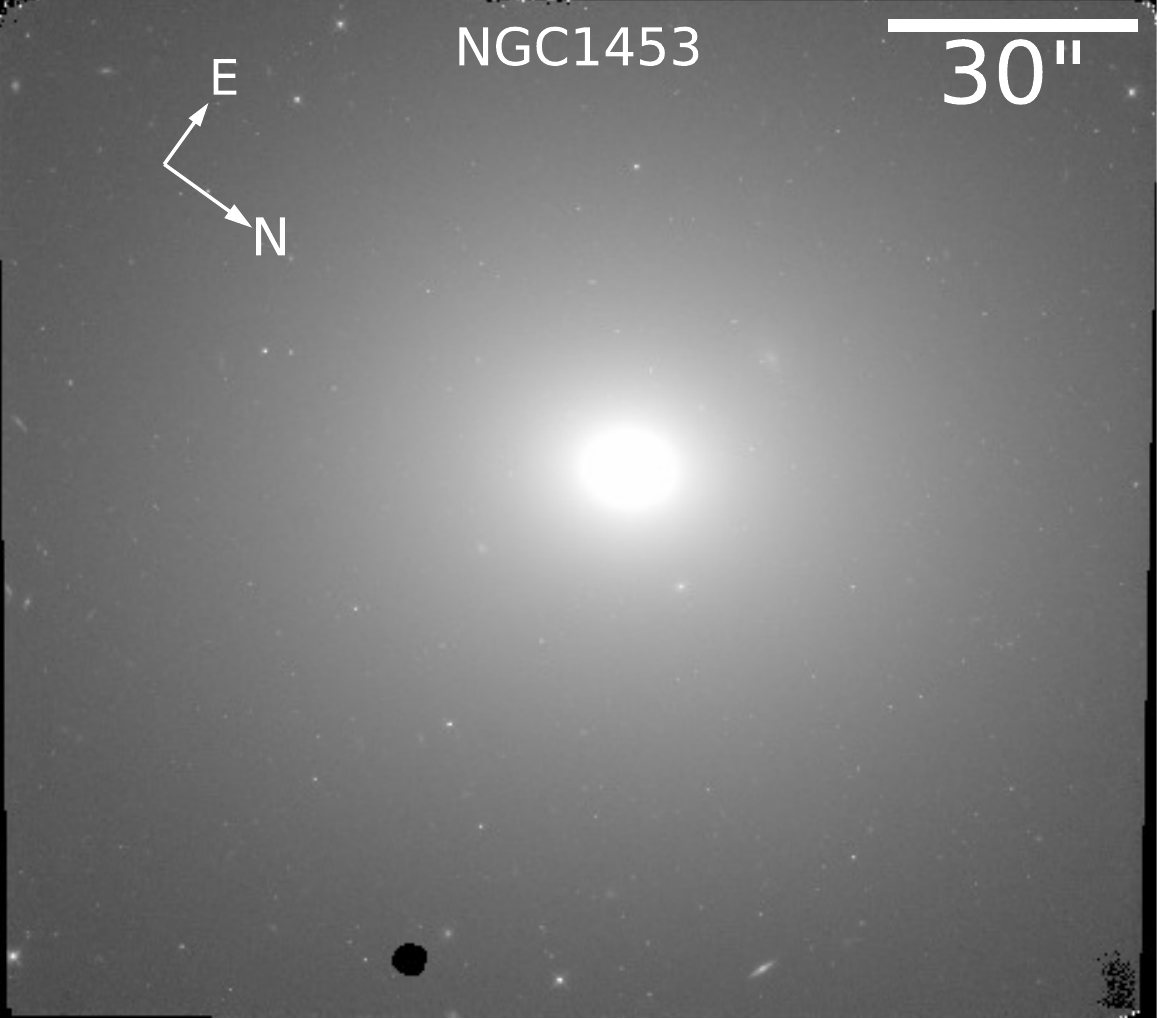}
	  \end{minipage} 
	  &     \begin{minipage}[t]{0.4\textwidth}
	    \vspace{0pt}\includegraphics[width=\textwidth]{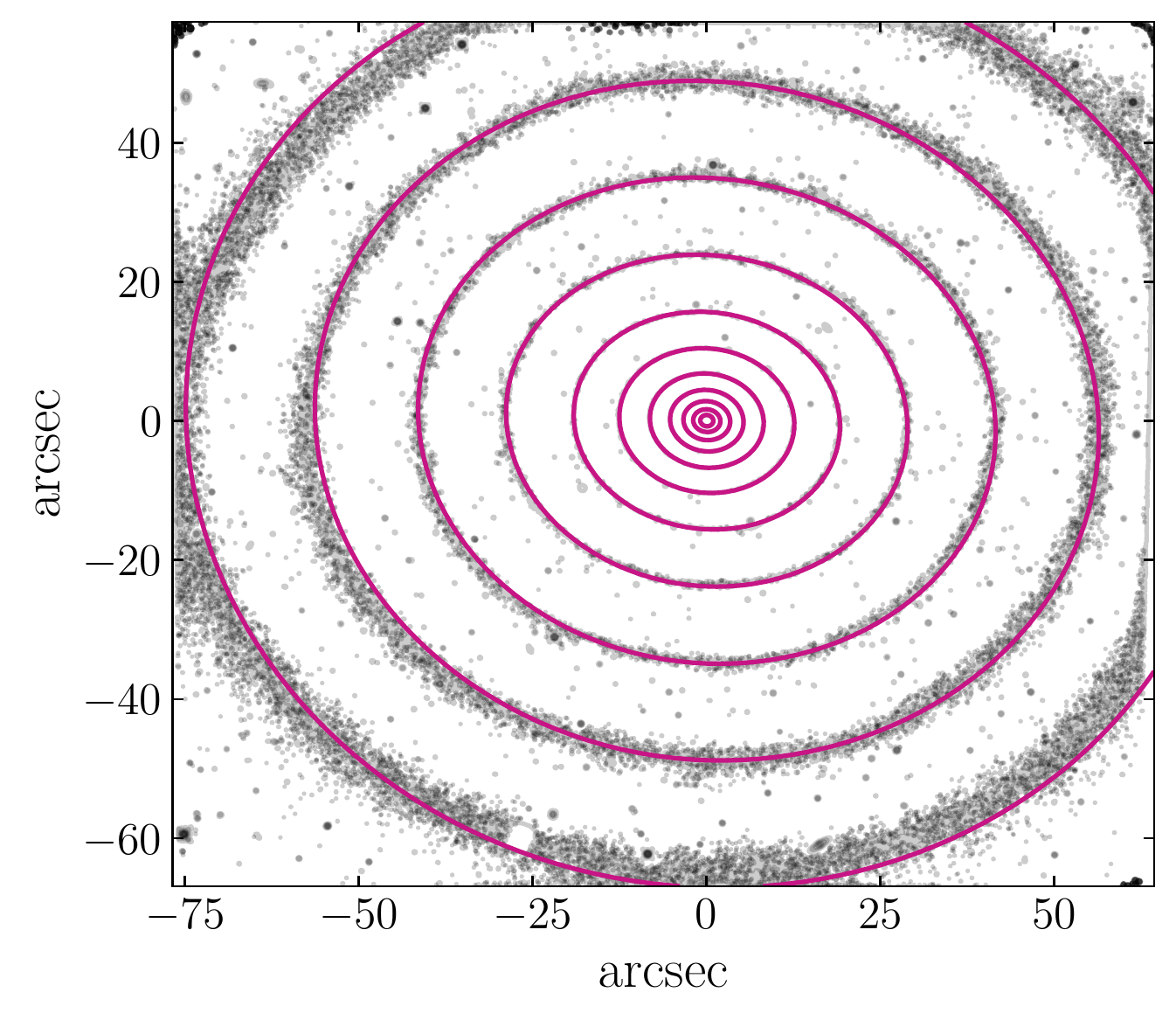}
	  \end{minipage}
	 \\
	 \includegraphics[width=0.45\textwidth]{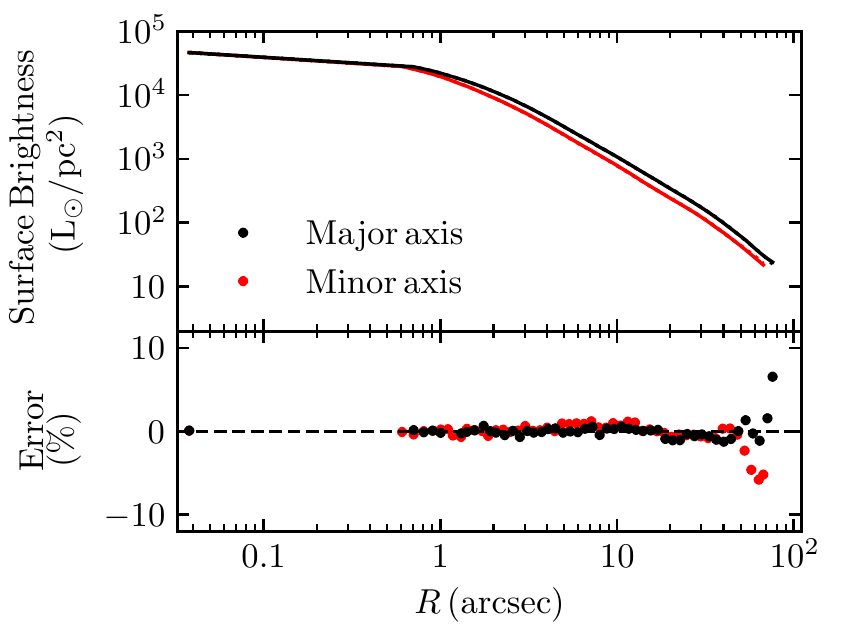} &   \includegraphics[width=0.45\textwidth]{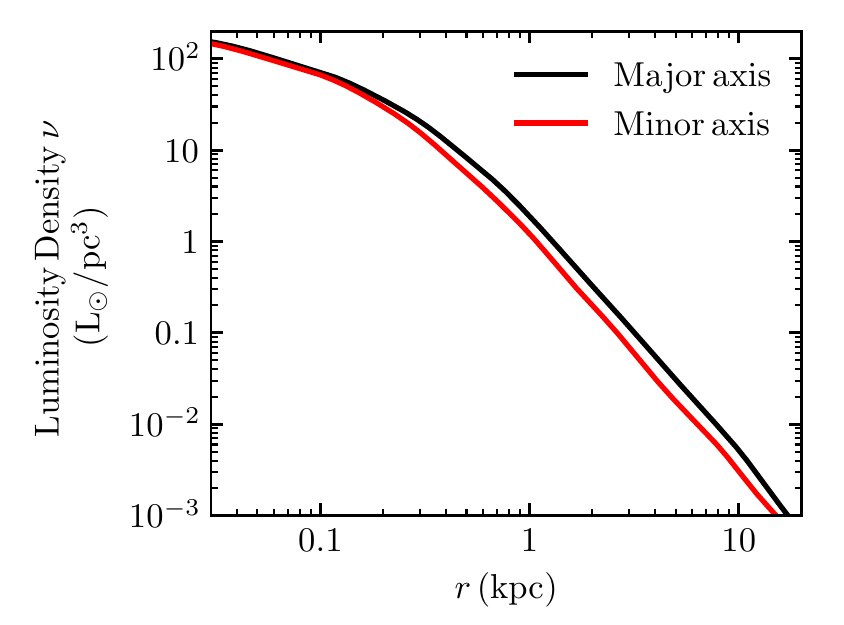} \\
	\end{tabular}

  \caption{(Upper left) The F110W-band HST image of \ngc\ used for our photometry \citep{Goullaudetal2018}.
  The image is oriented so that the $+x$ axis lies at $30.1^\circ$ east of north.
  (Upper right) Isophotes of the \emph{HST} WFC3 IR image of NGC~1453 (black) and the best-fit MGE model (magenta).  The isophotes have no measurable deviation from purely elliptical contours  \citep{Goullaudetal2018}. 
 (Lower left) The surface brightness profiles along the major (black) and minor (red) axes are well-fit by the sum of 10 Gaussians with small fitting errors. The difference between the data (solid) and model (dotted) is not discernible in the plot, where the fractional error (lower half of the panel) is $\sim 1$\% except at large radii beyond $50''$. (Lower right) Deprojected 3D luminosity density for an oblate axisymmetric model viewed edge-on for the best-fit MGE model.}
\label{fig:mge}
\end{figure*}

To model the spatial distribution of the stellar component of NGC~1453, we use the IR imaging portion of the MASSIVE survey with the F110W filter of the \emph{HST} Wide Field Camera 3 (WFC3/IR) \citep{Goullaudetal2018}.
The observations of \ngc\ had a total exposure time of 2496 seconds, which was divided into five dithered exposures using a five-point subpixel dither pattern to improve the point spread function (PSF) sampling.  
The pixel scale at F110W is 0.128 arcsec pix$^{-1}$ and is slightly undersampled for this wavelength. 
Details of the data reduction procedures, background sky measurement, mask construction, and isophotal fitting process were given in \citet{Goullaudetal2018}.

The isophotes of NGC~1453 are very regular (top panel of Fig.~\ref{fig:mge}) with a mean ellipticity of $0.17\pm 0.001$.
The position angle (PA) changes with radius mildly from $27.9^\circ\pm 1.0^\circ$ (east of north) at $1''$ to $36.1^\circ \pm 0.4^\circ$ at $79.5''$, with a luminosity-weighted average of $30.1^\circ\pm 0.2^\circ$. 
We fit the surface brightness using the Multi-Gaussian Expansion (MGE) method \citep{emsellemetal1994, Cappellari2002}
with a sum of 2D Gaussian components that share a common center and PA:
\begin{equation} \label{eg:mge}
    \Sigma (x^\prime, y^\prime) = \sum_{k=1}^{N} \frac{L_k}{2\pi \sigma^{\prime 2}_k q_k^\prime } {\rm exp} \Big[-\frac{1}{2 \sigma_k^{\prime 2} }\Big(x^{\prime 2} + \frac{y^{\prime 2}}{q_k^{\prime 2}}\Big)\Big],
\end{equation}
where $x^\prime$ and $y^\prime$ are projected coordinates measured from the galaxy center, with $x^\prime$ and $y^\prime$ being along the photometric major and minor axes, respectively. The subscript $k$ labels the individual Gaussian components; $L_k$, $\sigma_k^{\prime}$, $q_k^{\prime}$ are the luminosity, projected width, and projected axis ratio of each Gaussian, respectively. To compare to WFC3 images, we convolve the model with a PSF composed of 5 nearly-circular gaussian components (with axis ratios $>0.98$), obtained by fitting the PSF from \cite{Goullaudetal2018}.
The MGE fitting routine by default determines the PA using the central region of the galaxy. As a result, it chooses a PA of $28.5^\circ$, slightly different from the mean value $30.1^\circ$ quoted in \citet{Goullaudetal2018}. We repeated the MGE fit with the PA fixed to $30.1^\circ$ and found a virtually identical fit. We choose to use $28.5^\circ$, 
the value from the MGE fitting routine.

Our best-fit MGE to the surface brightness of NGC~1453 consists of 10 Gaussian components, which are summarized in Appendix~\ref{MGE_parameters} and plotted in Figure~\ref{fig:mge} (lower-left panel). The small fitting residuals (lower half of the panel) demonstrate that the MGE model agrees very well with the data.  This MGE fit has an effective radius $R_e = 19.6'' \approx 4.8$ kpc, very similar to $R_e=21.9''$ from \citet{Eneetal2018} using our deep K-band photometry from CFHT.

The intrinsic and projected coordinate systems are related by a set of three viewing angles $(\theta, \phi, \psi)$ \citep{Binney1985}.
The angles $\theta$ and $\phi$ specify how the line of sight is oriented relative to the principal axes of the galaxy, and $\psi$ specifies the rotation of the galaxy around the line of sight, where an oblate axisymmetric potential is defined to have $\psi=90^\circ$. Given these viewing angles, an MGE fit to the light profile $\Sigma(x^\prime,y^\prime)$ can be deprojected into a 3D luminosity density $\nu(x,y,z)$ with $x,y,z$ in the intrinsic coordinate system; see lower-right panel of Figure~\ref{fig:mge}.

Dust was not observed in the central region of \ngc\ in our WFC3 data.
The mean optical and UV colors of NGC1453 are typical of those of evolved, red giant ellipticals of similar masses (e.g., \citealt{Faberetal1989,Loubseretal2011}).
\citet{annibalietal2007} derived a mean age of 9.4 +/- 2.1 Gyr and metallicity [Z/H] = +0.22 dex within the central 3".  Thus, the dominant stellar population is old and metal-rich. We find no significant gradient in the g-z color from PanSTARRS data (Jensen et al. in prep).

\section{Schwarzschild Orbit Models} \label{Schwarzschild_Orbit_Models}

We use the orbit superposition method of \citet{Schwarzschild1979} through the implementation described by \citet{vandenBoschetal2008}. In this method, a library of orbits with a wide range of initial conditions is constructed
for a stationary potential due to a central black hole, a stellar component described by the MGE, and a dark matter halo. As each orbit passes through the region of the sky corresponding to a kinematic bin, its velocity is recorded to construct an LOSVD which is then decomposed in terms of Gauss-Hermite moments.  A superposition of orbits is constructed with the QPB quadratic programming solver from the GALAHAD library \citep{Gouldetal2003}, which minimizes the $\chi^2$ associated with the kinematics under the constraint that both the projected mass within each aperture and the 3D mass distribution are fit within 1\% of the MGE.

We have found several problems in the code during our tests and have fixed them as described in \matthewpaper. We have also determined that additional modifications are required to achieve axisymmetry within the code. These changes are discussed briefly in the following subsections and more fully in \matthewpaper.

\subsection{The axisymmetric limit} \label{axisymmetric_limit}

\ngc\ is a fast rotator with regular elliptical isophotes (Fig.~\ref{fig:mge}) and no significant misalignment between the projected rotation axis and photometric minor axis \citep{Eneetal2018}. 
These properties suggest that NGC 1453 can be approximated as an oblate axisymmetric model. 
However, we find the original version of the triaxial code by \citet{vandenBoschetal2008} not to be able to achieve exact axisymmetry.  
Here we describe two precautions and one change that we implemented in order to achieve axisymmetry.  

First, the box orbit library, which is generated by default in the original code, should be excluded when the code is to be used for axisymmetric gravitational potentials. Orbits in the box orbit library start from rest with $L_z=0$. These orbits are important in triaxial potentials, but not in axisymmetric systems where $L_z$ is an integral of motion. 
In an axisymmetric potential, box orbits cannot precess about the minor axis as they retain $L_z = 0$ for all time. As a result, they remain in their starting plane and do not exhibit axisymmetry.  We therefore exclude these intrinsically non-axisymmetric orbits from our axisymmetric models\footnote{For a triaxial potential, the ``loop'' orbit library generated by the code can contain some box orbits that have no overall sense of direction (see orbit start space in Fig.~2 of \citet{Schwarzschild1993}.  However, as $L_z$ is an integral of motion for an axisymmetric potential and all orbits in the library are initialized with $L_z \ne 0$, these orbits are also excluded.}.

The second precaution is to avoid generating long-axis tube orbits in the orbit library, a class of orbits not supported by axisymmetric stellar mass distributions. We find that these orbits can be eliminated only when the value for the viewing angle $\psi$ is set to be sufficiently close to $90^\circ$ in the input parameter file.
As described in Section~\ref{photometric_data}, the code uses three viewing angles $(\theta, \phi, \psi)$ to relate the intrinsic and projected coordinate systems and to set the axis ratios of the stellar potential. 
An oblate axisymmetric potential is obtained 
when $\psi$ is exactly $90^\circ$ and the axis ratio $p$ between the long and intermediate axes is exactly $1$.  Due to floating point imprecision, however,
the code does not run when $\psi$ is set to $90.0^\circ$ with double precision. Earlier work typically chose $|\psi - 90^\circ| = 10^{-3}$ or $10^{-2}$, assuming these values were close enough to $90^\circ$ to generate axisymmetry.
For \ngc, however, we find even $|\psi - 90^\circ| = 10^{-3}$ to be sufficiently far away from $90^\circ$ to allow for long-axis tubes in the orbit start space, hence violating axisymmetry.
We instead choose $\psi = (90 + 10^{-9})^\circ$ in this work, which is far enough from $90^\circ$ to avoid numerical issues but close enough that the potential is essentially axisymmetric for all available choices of the inclination.

Even after we excluded both the box and long-axis tube orbits,
we still were unable to achieve axisymmetry with the original triaxial code.  
In the case of \ngc, we find that many orbits precess on timescales much longer than the default integration time, which is set to be 200 times the orbital periods in the code.  These orbits should be symmeterized so that their contributions to the kinematics and mass grids are axisymmetric. To achieve this, we combine 40 copies of each orbit, each rotated slightly about the intrinsic minor axis of the galaxy; see \matthewpaper for details of our implementation.

These changes allow the triaxial code to be properly run in the axisymmetric limit.  We will compare results from the original code and our version in Section~\ref{BH_zero_solution}.

\subsection{Orbit library and phase space sampling}

As described in \citet{Schwarzschild1993}  and \citet{vandenBoschetal2008}, the orbits used for the models span a grid of energies ($E$) and starting positions ($R,\Theta$) on the meridional plane of the galaxy. We choose 40 energies corresponding to the potential energies $\Phi(r,0,0)$ evaluated at a set of 40 radii that are logarithmically spaced between 0.01 and $10^{2.5}\,\SI{}{arcsec}$. These radii are chosen to span from roughly one order of magnitude below the pixel scale of our photometry to the radii where $\ge 99.999\%$ of the MGE mass is contained.
We verify that orbits at the highest and lowest energies are given very low weight in the models.  Our tests also verify that adding orbits starting at higher or lower radii does not impact our models.  For each energy, we construct a grid of $9 \times 9$ starting positions spanning the radii between the inner and outer thin orbit radii for that energy and angles between $0^\circ$ and $90^\circ$.  

To improve the sampling of the phase space, the code allows orbit dithering where groups of orbits spanning a small volume in the $(E,R,\Theta)$ space are generated, combined, and given a single weight during orbit superposition. We use bundles of $5^3 = 125$ orbits for the final results below, and bundles of $3^3 = 27$ for numerous tests since they produce similar results and are less CPU-intensive.
Our models also include a time-reversed copy of each orbit.
In total, we use a library of 810,000 orbits (or 174,960 orbits for tests)
for each mass model
with $2 \times 40 \times 9 \times 9 = 6480$ independent weights.

As discussed in Section~\ref{axisymmetric_limit}, we use $\psi = (90 + 10^{-9})^\circ$ to run the triaxial code in the axisymmetric limit. In this limit, the $\phi$ viewing angle does not affect the MGE deprojection, but it sets the orientation of the plane of the orbit start space relative to the plane of the sky. As we axisymmetrize the orbits before projecting them onto the sky, the orientation between the start space and the sky becomes unimportant and we find that our model fits are independent of the viewing angle $\phi$. We choose $\phi = 1^\circ$.  For reference,
when the viewing angles $\psi = 90^\circ$ and $\theta = 90^\circ$ are used,  the choices of $\phi = 0^\circ$
and $\phi = 90^\circ$ correspond to aligning the intrinsic $x$-axis and $y$-axis with the line of sight, respectively.

The potential due to the central black hole includes a softening length so that the potential at the origin is not singular. We set this length to $\SI{3e-4}{arcsec}$, which is roughly two and a half orders of magnitude smaller than the size of our central kinematic bin and one and a half orders of magnitude smaller than the peribothron of the most central orbits. 

We convolve the integrated orbit trajectories in the models with PSFs while projecting the orbits onto the sky. This convolution is done separately for each kinematic dataset as they have different PSFs. For each, we assume a single circularly symmetric Gaussian with a FWHM of $0.7''$ for the GMOS kinematics and $1.2''$ for the Mitchell kinematics.

\subsection{Input Gauss-Hermite moments}
\label{losvd_fitting}

We use the first 12 moments in the Gauss-Hermite expansion of the LOSVDs as constraints in the orbit models.  
For the central region of \ngc, we use the first 8 moments $V$, $\sigma$, $h_3$, \dots, $h_8$ measured from the GMOS spectra as described in Section~\ref{central_kpc_kinematics} and shown in Figure~\ref{fig:gmos_map}.  
The corresponding radial profile of each of the moments for all 135 GMOS spatial bins is plotted (black points) in the left panel of Figure~\ref{figure:kinematics_radial}. 
The errors on $h_3$ through $h_8$ are quite similar from moment to moment and bin to bin.  The mean errors on these moments range from 0.018 to 0.023, with a typical standard deviation of 0.003 over the spatial bins.
To choose an appropriate number of moments to extract using pPXF, we performed the extraction with increasing numbers of moments (4, 6, 8). As the number of extracted moments is increased, we find that the typical value of the highest extracted moment becomes consistent with 0. For the GMOS spectra, this occured when 8 moments were extracted.

To prevent spurious behavior in the higher-order moments in the model, we further constrain the next four orders, $h_9$ to $h_{12}$, to be $0.0 \pm \delta$, where $\delta$ represents the typical errors in the higher moments. 
Since the size of errors is very similar from $h_3$ to $h_8$, we do not find the exact assigned values of $\delta$ to matter. Nonetheless, we try to mimic the mild bin-to-bin variations by assigning
the measured errors for $h_7$ for a given bin to $\delta$ for the odd moments $h_9$ and $h_{11}$ in that bin, and similarly
for the even moments (i.e., using
the $h_8$ errors for $h_{10}$ and $h_{12}$).

For the wide-field data that have lower $S/N$, we use the first 6 Gauss-Hermite moments measured from the Mitchell spectra as constraints (Sec.~\ref{wide_field_kinematics}).
The radial profile of the moments for the 38 Mitchell spatial bins extending to a radius of $\sim 50''$ is shown in the right panel of Figure~\ref{figure:kinematics_radial}. 
We again constrain the 7th and 12th moments to be 0 with uncertainties equal to the measured errors for $h_5$ (for odd orders) or $h_6$ (for even orders).
The errors on moments $h_3$ through $h_6$ from the Mitchell spectra are also quite uniform between moments. The mean errors on these moments range from 0.029 to 0.035 with a typical standard deviation of 0.006 over the spatial bins.

We discuss further the importance of constraining the higher Gauss-Hermite moments in Sec.~7.1 below.

\section{Results: Mass Model Search}

\subsection{Mass Model}

We investigate four mass model parameters -- inclination $\theta$, central black hole mass \mbh, F110W-band stellar mass-to-light \ml, and the enclosed mass of the dark matter halo at $15$ kpc.
We use a logarithmic halo with mass density
\begin{equation} \label{eq:rho_dm}
\rho_{DM}(r) = \frac{V_c^2}{4 \pi G} \frac{3 R_c^2 + r^2}{(R_c^2 + r^2)^2}  \,.
\end{equation}
We find the circular velocity $V_c$ and the scale radius $R_c$
to be highly degenerate for our data because
the enclosed mass
\begin{equation} \label{eq:encmass_dm}
M_{enc}(r) = \frac{V_c^2}{G} \frac{r^3}{r^2 + R_c^2}
\end{equation}
scales with $V_c^2 / R_c^2$ within the scale radius where most of our data points are located.  
We therefore choose to parameterize the halo with the enclosed mass within 15 kpc, $M_{15}$,  where
15 kpc is the middle of the radial extent of the outermost Mitchell bins (spanning 9.4 kpc to 18.8 kpc).

\subsection{Marginalization}

Previous orbit modeling papers
have often determined the $1\sigma$ ($68\%$) and $3\sigma$ ($95\%$) confidence intervals for each model parameter by finding the values at which the $\chi^2$ rises by $\Delta \chi^2 = 1$ and $9$ relative to the best-fit model.  This method is only exactly correct when there is no covariance between the marginalized and free parameters and where the free parameter's $\chi^2$ landscape is quadratic so that the likelihood is Gaussian. To avoid reliance on these assumptions, we compute best-fit values and confidence intervals through an interpolation and marginalization routine described in Appendix~\ref{interpolation}.

\begin{figure}
  \centering
  \includegraphics[width=0.95\linewidth]{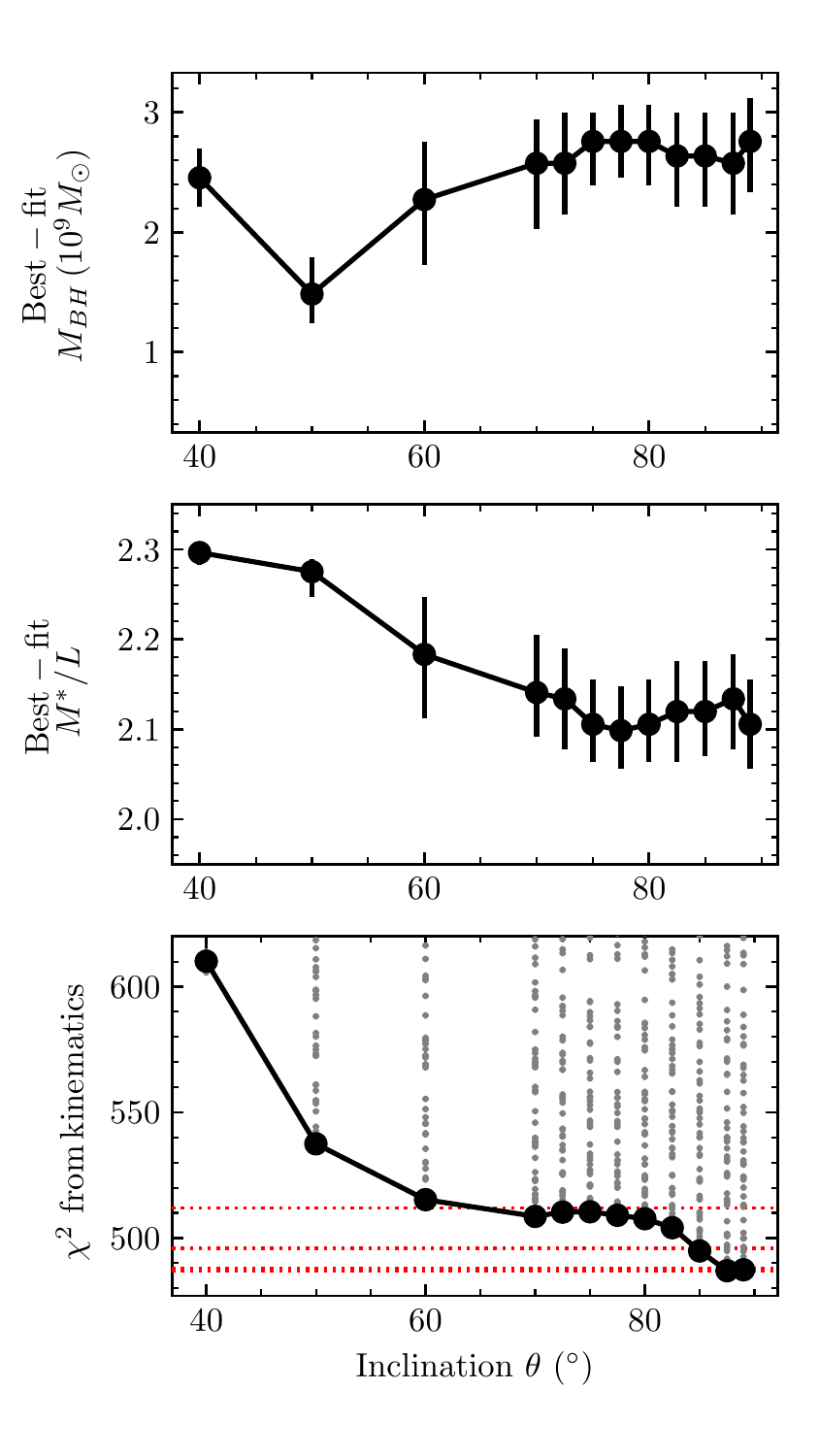}
  \hfill
  \caption{ Best-fit \mbh\ (top) and \ml\ (middle) with $1\sigma$ confidence intervals and the corresponding marginalized 1-d $\chi^2$ (bottom) as a function of the inclination angle $\theta$.
  The grey points in the lower panel denote the $\chi^2$ of individual models within the grid, and the red horizontal dashed lines denote the conventional $0,1,3,5\sigma$ confidence levels corresponding to $\Delta \chi^2 = 0,1,9,25$. 
  The halo is fixed to $M_{15} = \SI{6.3e11}{M_\odot}$ in this plot for illustrative purposes; similar dependence is found for other halo masses we examined.
  \label{figure:1DvsTheta}}
\end{figure}

\subsection{Inclination}

In the oblate axisymmetric limit ($p = 1$), the MGE deprojection requires a single viewing angle, $\theta$, which specifies the assumed inclination of the galaxy. An edge-on view of the system corresponds to $\theta = 90^\circ$ while face-on corresponds to $\theta = 0^\circ$. The inclination affects the axis ratios of the deprojected density distribution with $q_i = \sqrt{q^{\prime 2}_i - \cos^2{\theta}}/\sin \theta$, where $q'_i$ is the observed axis ratio of the $i-$th component of the MGE fit described in Appendix~\ref{MGE_parameters}, and $q_i$ is the intrinsic axis ratio between short axis to the long axis in that component's deprojection. 

Deprojection is only possible when $\cos^{-1} (\min{q'_i}) < \theta < 90^\circ$, where $\min{q'_i}$ is the smallest axis ratio in the MGE fit. For the MGE used in this analysis, we have $\cos^{-1} (\min{q'_i}) = 38^\circ$.  When inclinations near this threshold are used, flattening of the the MGE component with the smallest $q'$ changes significantly. For example, when $\theta = 40^\circ$, the component with $q' = 0.786$ has $q = 0.27$, and for $\theta = 50^\circ$, the component is flattened to $q = 0.59$.

To determine the inclination of NGC1453, we search coarsely over $\mbh$, $\ml$, and $\mdm$ but finely over the inclination. 
This grid includes 11 values of \mbh\ from $0.0$ to $\SI{6e9}{}M_\odot$ in steps of $\SI{6e8}{}M_\odot$, 8 values of \ml\ from $1.6$ to $2.3$ (in solar units) in steps of $0.1$, and 3 values of $\mdm$: $2.8$, $6.3$, and $11.2$ $\times 10^{11} M_\odot$, corresponding to $V_c = 400$, $600$, and $\SI{800}{\kms}$ with $R_c = \SI{15}{kpc}$. We use 12 values of $\theta$ from $ 40^\circ$ to $89^\circ$ in steps of $10^\circ$ below $70^\circ$ and $2.5^\circ$ above. The code does not allow perfectly edge-on viewing angles, so the highest $\theta$ sampled was $89^\circ$ rather than $90^\circ$. This grid contains $11 \times 8 \times 3 = 264$ models for each choice of $\theta$ and $264 \times 12 = 3168$ models in total.

Figure~\ref{figure:1DvsTheta} shows that nearly edge-on viewing angles are strongly preferred.  When $\mdm = \SI{6.3e11}{M_\odot}$, $\theta = 89^\circ$ gave the lowest overall $\chi^2$ with $\theta = 80^\circ$ and $70^\circ$ lying $\Delta \chi^2 = 20.9$ and $22.3$ higher. 
For each $\theta$ and $\mdm$, we compute the best-fit \mbh\ and \ml\ with their $68\%$ confidence intervals. When $\theta > 70^\circ$, the best-fit values depend only weakly on $\theta$, and their confidence intervals coincide. This suggests that our recovered black hole mass and stellar mass-to-light ratio are relatively insensitive to the inclination within the edge-on limit.  We therefore fix the inclination to be nearly edge on with $\theta = 89^\circ$ as we sample over halos below.

\subsection{Black hole, stars, and dark matter halo}\label{Vc_grid}

With the inclination fixed to be nearly edge-on with $\theta = 89^\circ$, we search the three mass parameters, \mbh, \ml, and $\mdm$, using two sets of grids.  The primary grid covers the parameter ranges broadly and is then supplemented by a finer grid that zooms into the best-fit model of the primary grid with half the grid spacing in both \mbh\ and \ml.  

The primary grid has $16\times 15\times 13=3120$ models for \mbh, \ml\ and $\mdm$. 
This grid samples \mbh\ linearly from 0 to $\SI{6e9}{} \msun$ in steps of $\SI{4e8}{} \msun$,  $\ml$ from 1.60 to 2.30 (in solar units) in equal steps of 0.05, and the enclosed halo mass from $\mdm = \SI{2.79e11}{M_\odot}$ to $\SI{11.16e11}{M_\odot}$ by varying the circular velocity roughly linearly from $\vc = 400 \kms$ to $800 \kms$ (for $R_c=15$ kpc).

For the finer grid, we first determine the \mbh\ and \ml\ model that minimizes the $\chi^2$ for each value of $\mdm$.  We then construct the fine grid around that model sampling another $16 \times 15$ values of $\mbh$ and $\ml$, where the spacing between models is half of that of the primary grid, and $\mbh$ is sampled over a range of $\SI{3e9}{} \msun$ in steps of $\SI{2e8}{} \msun$, and $\ml$ is sampled over a range of 0.35 in steps of 0.025. Many of these models overlap with those of the primary grid, so only $176 \times 13 = 2288$ additional models are run. 

\begin{figure}
  \centering
\includegraphics[width=3.7in]{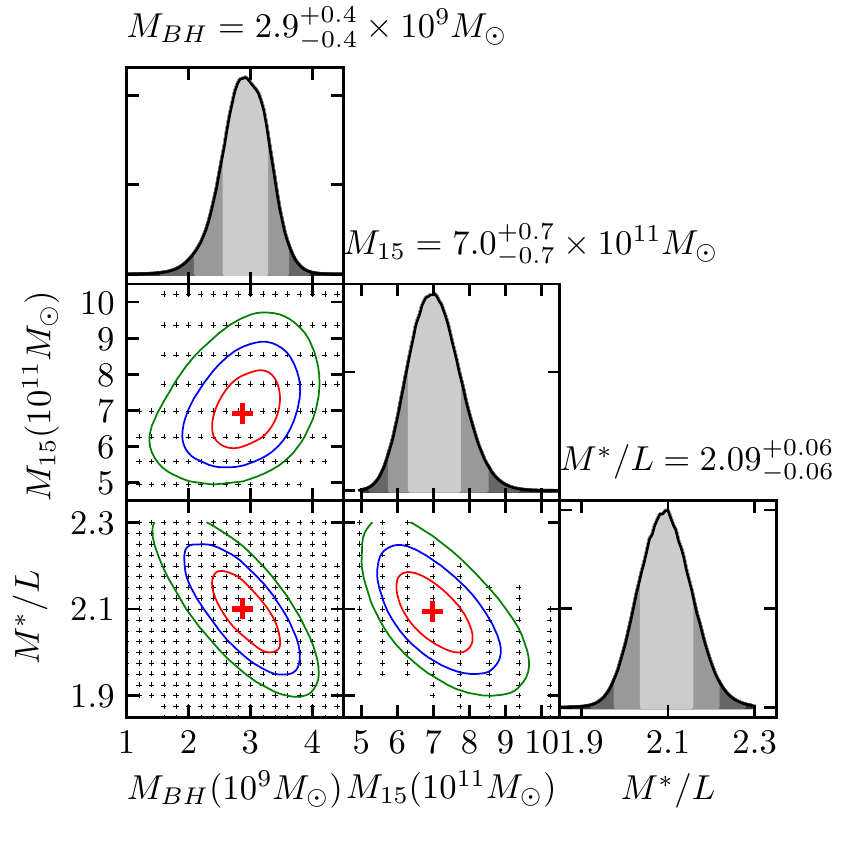}
  \hfill
  \caption{Marginalized 1-d and 2-d likelihood distributions from the grids of \mbh, $\mdm$ and $\ml$ described in Section.~\ref{Vc_grid}. 
  The $1\sigma$, $2\sigma$, and $3\sigma$ confidence intervals corresponding to the 68, 97, and 99.5 percentile confidence levels are shown as red, blue, and green curves in the 2-d panels and as different shade of grey in the 1-d panels. 
  The extracted best-fit values and $1\sigma$ confidence interval are shown above each 1-d panel.
  \label{figure:CornerPlots}}
\end{figure}

\begin{figure}[h]
  \centering
\includegraphics[width=\linewidth]{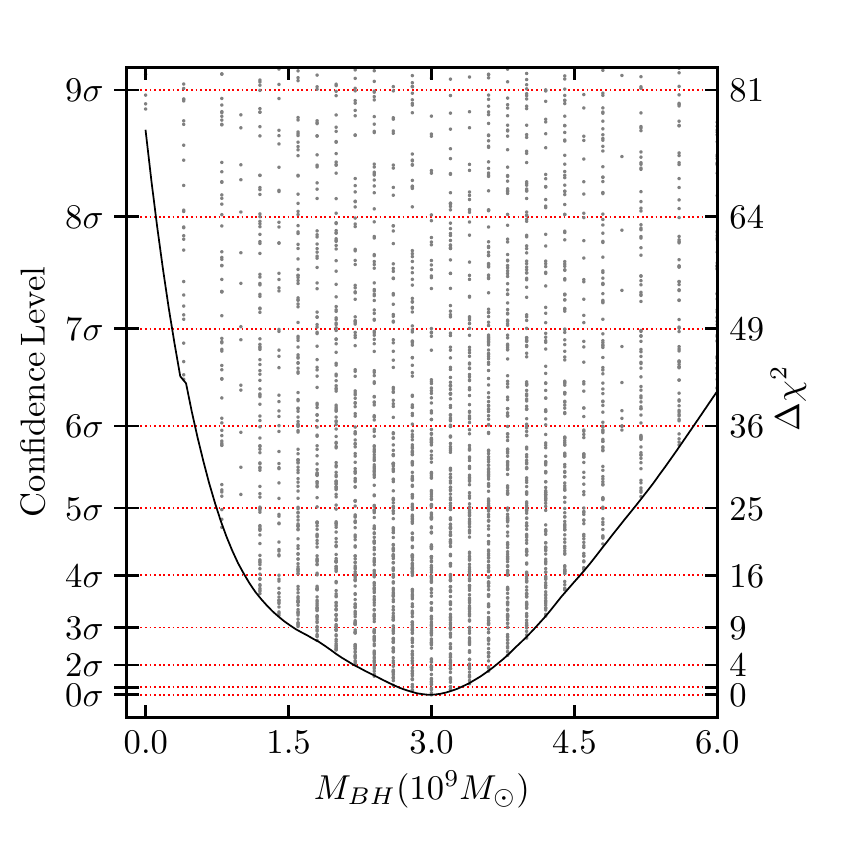}
  \hfill
  \caption{ $\chi^2$ as a function of \mbh\ for the full range of \mbh\ explored in this paper.
  The $\chi^2$ is obtained by marginalizing over the other two mass parameters, $\mdm$ and $\ml$, as described in Appendix~\ref{interpolation}.  Models with $\mbh = 0$ are highly disfavored with a $\Delta \chi^2 = 75.5$ relative to the best-fit $\mbh$, corresponding to the $8.7\sigma$ confidence level. 
  \label{figure:ChisqVsMbh}}
\end{figure}

We perform the interpolation and marginalization described in Appendix~\ref{interpolation} to determine the best-fit values and uncertainties in \mbh, \ml\ and $\mdm$ from these 5408 models.
The resulting $\chi^2$ landscapes are displayed in Figure~\ref{figure:CornerPlots}. We find the best-fit mass parameters to be $\mbh = (2.9\pm 0.4)\times 10^9 M_\odot$, $\ml = (2.09\pm 0.06) M_{\odot} / L_{\odot}$, and $\mdm = (7.0\pm0.7)\times 10^{11} M_\odot $.
For comparison, if the best-fit parameters are chosen by finding the range of models where the $\chi^2$ rises by $\Delta \chi^2 \le 1$ from the minimum value, as was frequently done in prior \mbh\ papers, we find comparable central values for the mass parameters for \ngc, but the error bars are underestimated by a factor of 1.5 to 2: $\mbh = (3.0\pm 0.2)\times 10^9 M_\odot$, $\ml = (2.06\pm 0.03) M_{\odot} / L_{\odot}$, and $\mdm = (7.4\pm 0.4)\times 10^{11} M_\odot $. These values are tabulated in Table~\ref{tab:bestFitMinVsMarg}.

Figure~\ref{figure:ChisqVsMbh} presents a clear view of the $\chi^2$ landscape over the wide range of \mbh\ covered by our grid,   
It shows that models with small black hole masses are highly disfavored. In particular, $M_{BH} = 0$ has $\Delta \chi^2 = 75.5$ above the minimum, corresponding to the $8.7\sigma$ confidence level. This result will be further discussed in Section~\ref{BH_zero_solution}. 

In the best-fit mass model for \ngc, the enclosed stellar mass is equal to \mbh, 2\mbh, 3\mbh\ and 5\mbh\ at radius 0.18 kpc ($0.74''$),  0.26 kpc ($1.05''$), 0.33 kpc ($1.32''$), and 0.45 kpc ($1.83''$), respectively.
The enclosed stellar mass equals that of the dark matter at 8.4 kpc ($34.1''$).  At the effective radius (5 kpc), the dark matter fraction is 0.27.

\begin{table}
\centering

\caption{Best-fit values of the black hole mass, stellar mass-to-light in the F110W band, and dark matter mass enclosed within $15$ kpc. The center column presents values determined though interpolation and marginalization as described in Appendix~\ref{interpolation}. The right column presents values determined through projection, where the confidence interval bounds all models within $\Delta \chi^2 \le 1$ of the global minimum.
\label{tab:bestFitMinVsMarg}}

\begin{tabular}{ccc}
Mass parameters  & Marginalized & Projected  \\
\hline
$\mbh \,(10^9 M_\odot) $ & $ 2.9 \pm 0.4 $ & $3.0 \pm 0.2$  \\
$\ml \,(M_\odot / L_\odot)$  &  $ 2.09 \pm 0.06 $ &  $ 2.06\pm 0.03$             \\
$\mdm  \,(10^{11} M_\odot)$ &  $ 7.0\pm0.7 $
&  $ 7.4\pm0.4$          \\
\hline
\end{tabular}

\end{table}

\section{Results: Best-Fit Mass Model}\label{Bestfit_model}

\subsection{Stellar kinematics}

Our best-fit mass model (red points in Fig.~\ref{figure:kinematics_radial}) provides an excellent fit to the observed stellar kinematics (black).
Both the rotation $V$ and the large central $\sigma$ are well captured by this model.
The total $\chi^2$ for the best-fit model from all the kinematic moments is 493.0, where the bulk of this (471.5)
comes from the moments extracted from data and only a small fraction (21.5) comes from the additional high moments that are constrained to be zero.

To estimate the reduced $\chi^2$, we note that there are 8 measured moments for each of the 135 GMOS bins and 6 measured moments for each of the 38 Mitchell bins, for a total of 1308 data points. The kinematic maps of the odd moments have been point-anti-symmetrized and the even moments have been point-symmetrized according to the prescription in Appendix A of \citet{vandenboschdezeeuw2010}. 
Our total reduced $\chi^2$  from the moments extracted from data is therefore $471.5 / 1308 = 0.36$. The reduced $\chi^2$ from the GMOS data alone is $366.2 / (135 * 8) = 0.34$ and the reduced $\chi^2$ from Mitchell alone is $105.3 / (38 * 6) = 0.46$.  The high moments which were constrained to be zero have an associated reduced $\chi^2$ of $21.5 / 768 = 0.03$.

\subsection{Orbital structure}\label{Orbital_structure}

While computing the orbit libraries, the code constructs a 3D spherical grid containing the first and second velocity moments of the orbits. We use this velocity grid to compute the anisotropy parameter $\beta = 1 - \sigma_t^2 / \sigma_r^2$ and the ratio of radial to tangential dispersions $\sigma_r / \sigma_t$.  We note that various definitions of $\beta$ have been adopted in prior papers, and at times it is unclear whether $\sigma$ in such quantities is treated as a dispersion or a second moment of the velocity, i.e., whether $\sigma^2 = \langle v^2 \rangle - \langle v \rangle^2$ or $\sigma^2 = \langle v^2 \rangle$.  We choose to define
\begin{align*}
\sigma_{t}^2   &= \frac{\sigma_\theta^2 + \sigma_\phi^2}{2}        \,,       &   
& \beta = 1 - \frac{\sigma_{t}^2}{\sigma_r^2} \,,\\
\sigma_{rot}^2 &= \frac{\sigma_\theta^2 + \sigma_\phi^2 + \langle v_\phi \rangle^2}{2} \,,  &  & \beta_{rot} = 1 - \frac{\sigma_{rot}^2}{\sigma_r^2} \,,
\end{align*}
where the brackets denote a mass-weighted mean over $\theta$ and $\phi$. These pairs of definitions are only expected to differ when there is significant contribution from the ordered flow velocity term $\langle v_\phi \rangle^2$.
For reference, differing definitions and symbols were used in the literature, e.g., $\beta$ from \citet{Thomasetal2014}, $\beta_r$ from \citet{Peletieretal2007}, and $\sigma_r / \sigma_t$ from \citet{Walshetal2015} all excluded the $\langle v_\phi \rangle^2$ term, while $\beta_{rot}$ from \citet{Krajnovicetal2018} and \citet{Thomasetal2014} and $\sigma_{r} / \sigma_t$ from \citet{Gebhardtetal2003} included this term.

\begin{figure}
  \centering
  \includegraphics[width=\linewidth]{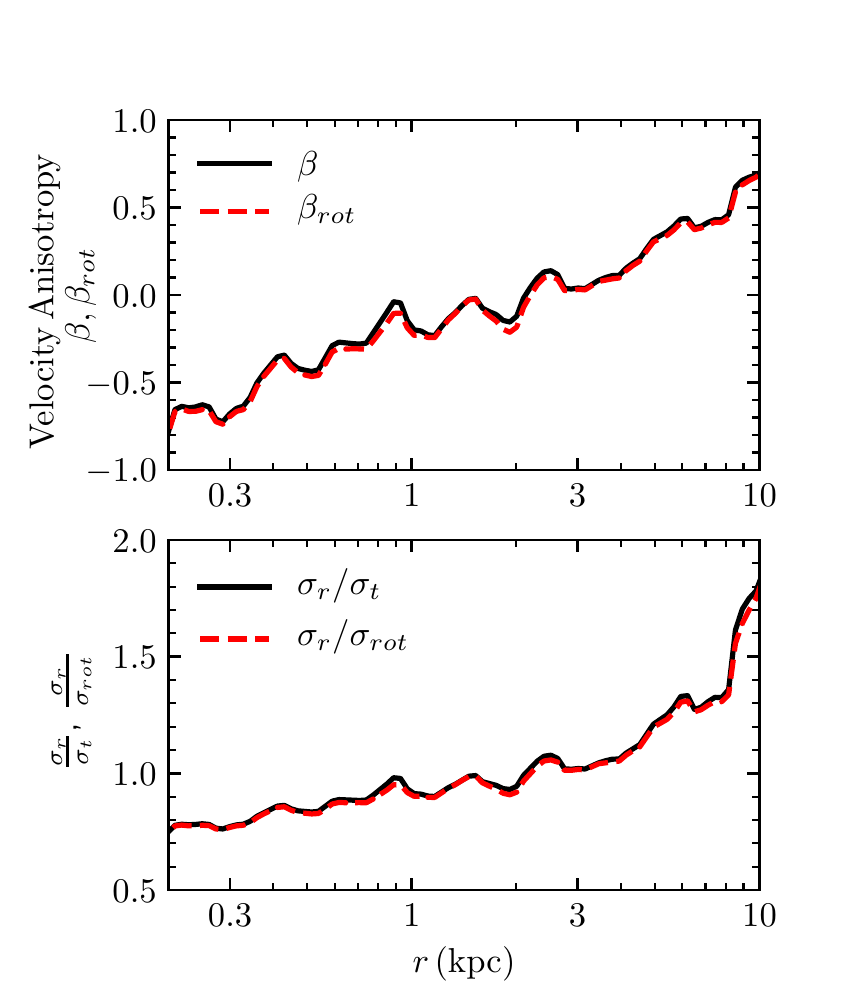}
  \hfill
  \caption{ Velocity anisotropy $\beta$ (top) and $\sigma_r /\sigma_t$ (bottom) as a function of radius for the best-fit model of \ngc.
   The orbits within the central $\sim 1$ kpc are preferentially tangential with $\sigma_r /\sigma_t < 1$ and $\beta < 0$.
   The orbits become increasingly radial beyond the effective radius ($\approx 5$ kpc)
   Over all radii, $\beta$ traces $\beta_{rot}$ and $\sigma_t$ traces $\sigma_{rot}$ because $\langle v_\phi \rangle^2 \ll \sigma_\phi^2$.
 }
\label{figure:Betas}
\end{figure}

The resulting velocity anisotropy as a function of spherical radius $r$ for the best-fit model of \ngc\ is shown in Figure~\ref{figure:Betas}.  The orbits are tangential near the core, but become increasingly radially anisotropic beyond the effective radius ($\sim 5$ kpc)  Even though \ngc\ exhibits rotation and is considered a fast rotator for an ETG,
the maximal velocity observed in our kinematics is $\sim \SI{100}{\kilo\meter\per\second}$, which is much below the dispersion $\sigma$ shown in Figure~\ref{figure:kinematics_radial}.
The term $\langle v_\phi \rangle^2$ therefore has negligible impact on the value of $\sigma_{rot}$ and $\beta_{rot}$ at all radii, and
$\sigma_{rot} \approx \sigma_{t}$ and $\beta \approx \beta_{rot}$ at all radii. 

\citet{Thomasetal2014} studied eleven massive elliptical galaxies with axisymmetric Schwarzschild models. Six of those galaxies had stellar cores and exhbited strongly tangential anisotropies ($\beta_{rot} < -0.5$) in the core regions and highly radial anisotropies ($\beta_{rot} \sim 0.5$) well outside the cores. Similar trends in the anisotropy were found in MASSIVE survey galaxy NGC~1600 \citep{Thomasetal2016}. This behavior is consistent with gravitational core scouring, where a central binary black hole preferentially ejects radial orbits from the core leaving an orbital structure which is tangentially biased \citep{Begelmanetal1980}. We observe similar behavior in \ngc\, suggesting that its core may have also been depleted through core-scouring.

\section{Discussion}

\subsection{Gauss-Hermite series truncation and LOSVDs}\label{GH_discussion}

As described above, the stellar LOSVD in each spatial bin is parameterized by a Gauss-Hermite series up to order $n$.  Some care must be taken to ensure 
that the unconstrained higher moments beyond order $n$ in the orbit models do not introduce spurious behavior in the predicted LOSVDs.

\begin{figure*}
  \centering
  \includegraphics[width=\linewidth]{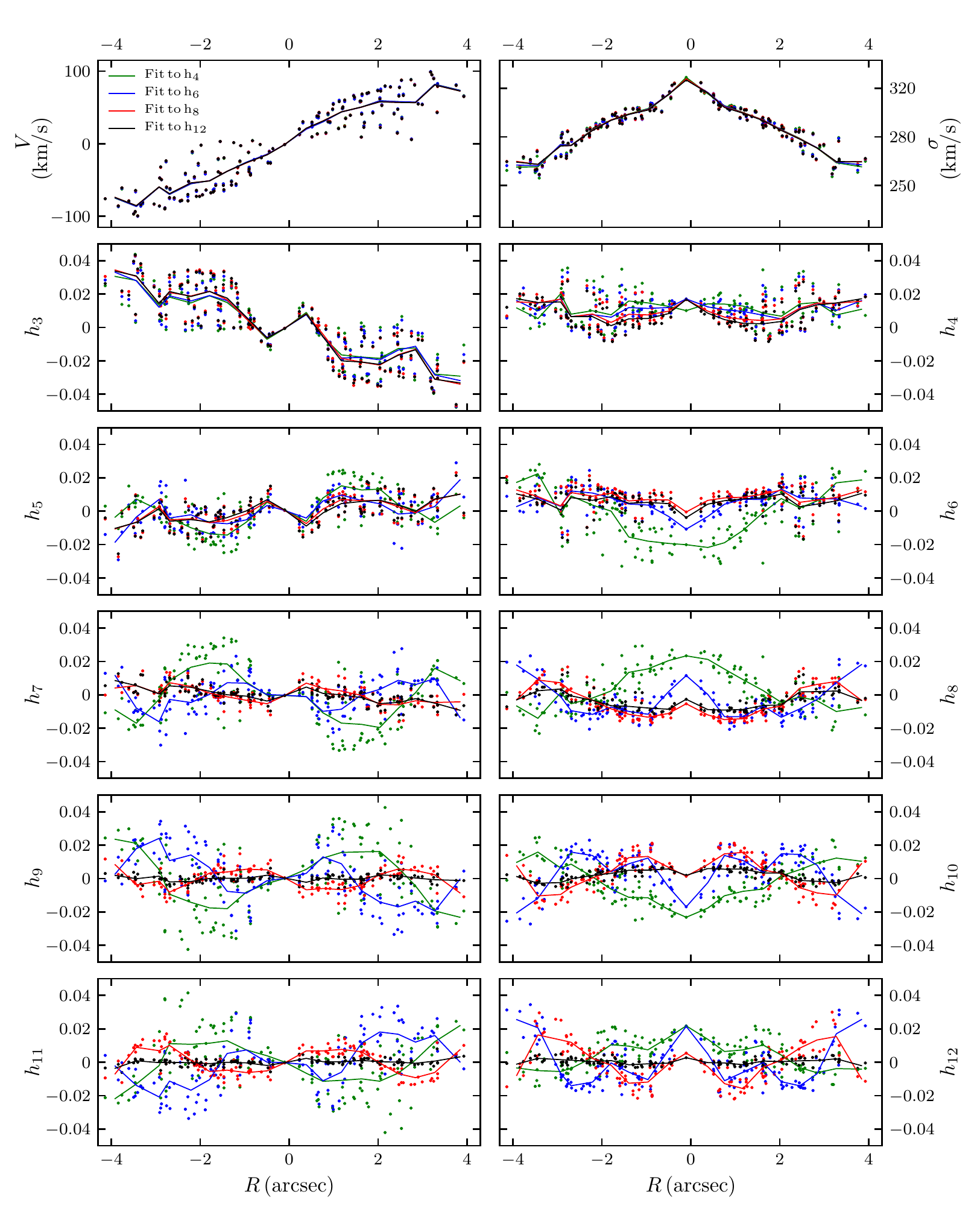}
  \hfill
  \caption{Dependence of the first 12 Gauss-Hermite moments predicted by the best-fit orbit model on the assumed truncation order applied to the GMOS data and used as input constraints.  
  The four colors show the progression of increasing truncation order: up to $h_4$ (green), $h_6$ (blue), $h_8$ (red), and $h_{12}$ (black; our production run).   In each case, the moments beyond the truncation order are unconstrained in the orbit model and exhibit correlated deviations from 0. This is most clearly seen in the green curves in the unconstrained $h_5$ and above. An interpolating line has been added to each curve to guide the eye. 
  The corresponding marginalized $\chi^2$ versus \mbh\ for the four cases are shown in Figure~12(left panel).
  See text in Sec.~7.1 for details.
  \label{figure:higher_moments}}
\end{figure*}

\begin{figure}
  \centering
  \includegraphics[width=\linewidth]{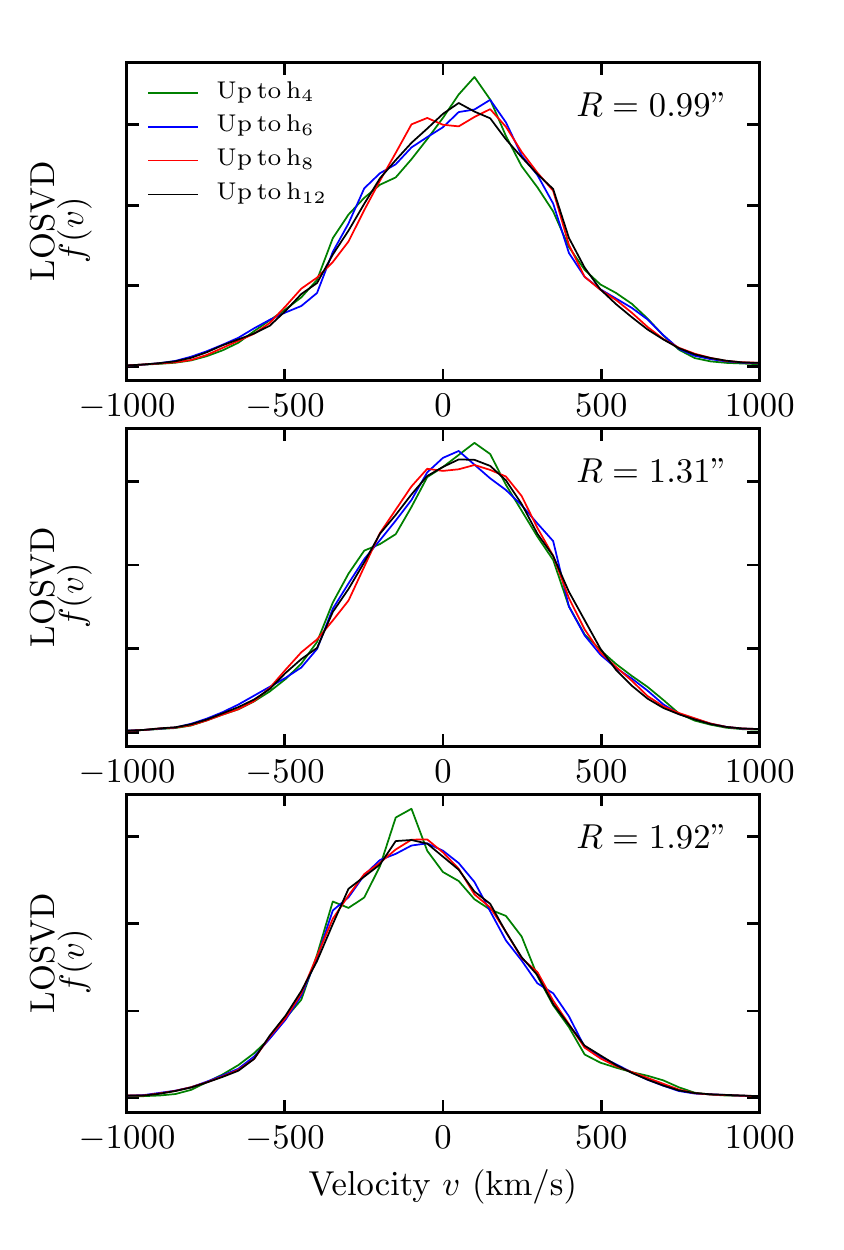}
  \hfill
\caption{LOSVDs for three representative GMOS bins predicted by the best-fit orbit models. Each panel 
compares the LOSVDs from the four models described in Sec.~\ref{GH_discussion} and shown in Figure~\ref{figure:higher_moments}, where the number of Gauss-Hermite moments fit by the dynamical models varies from 4 (green), 6 (blue), 8 (red), to 12 (black). When only 4 moments are constrained, the LOSVDs have the most pronounced irregular features due to the unconstrained $h_5$ and beyond seen in Figure~\ref{figure:higher_moments}. These unobserved features are gradually reduced when higher moments are used to constrain the model. 
 \label{figure:LOSVDs}}
\end{figure}

It is useful to begin the discussion by examining how the LOSVDs are obtained in the triaxial orbit code.  During orbit integration, the code first computes the LOSVD of each orbit for a spatial bin as it passes through the aperture on the sky. The Gauss-Hermite moments are then determined from each LOSVD through the direct integration described by \citet{vanderMarelFranx1993}, using the observed $V$ and $\sigma$ values for that bin.  During the subsequent orbital weight finding process, the Gauss-Hermite moments of the superposition of orbits in each bin
are easily computed due to their linearity.

To find the best-fit orbital weights for a mass model, the code places no constraints on moments beyond those provided to it. We are concerned that  some of the unconstrained higher moments can 
add spurious non-Gaussian features to the LOSVDs.
This is to be contrasted with how the LOSVDs are determined from the data. There, the pPXF algorithm determines the best-fit moments from the observed spectrum in the least-squares sense, choosing moments that minimize the residual contribution from higher moments.  

To test the impact of unconstrained higher moments on \mbh, we perform a series of controlled experiments in which we vary systematically
the number of Gauss-Hermite moments determined from the GMOS spectra and used as constraints in the orbit model. 
We compare the results from grid searches for four cases here. For the first three cases, 4, 6, and 8 GH moments are measured from the GMOS data with pPXF and those 4, 6, and 8 GH moments are fit with orbit models to infer \mbh. In the fourth case, 8 GH moments are determined from the GMOS data with pPXF and 12 GH moments are used as inputs into the dynamical models, with the 9th to 12th moments set to 0 and assigned uncertainties as described in Sec.~\ref{losvd_fitting}. The fourth case, where 12 moments are used to constrain the dynamical models, corresponds to our production run reported in earlier sections.
For each case, we then perform a grid search for the best-fit \mbh\ and \ml.
Our aim here is to test the effects on the measured \mbh, so we keep the large-scale Mitchell kinematics unchanged and fix the halo to the best-fit value of $\mdm =7\times 10^{11} M_\odot$ from our production run. 

\begin{figure*}
  \centering
\includegraphics[width=\textwidth]{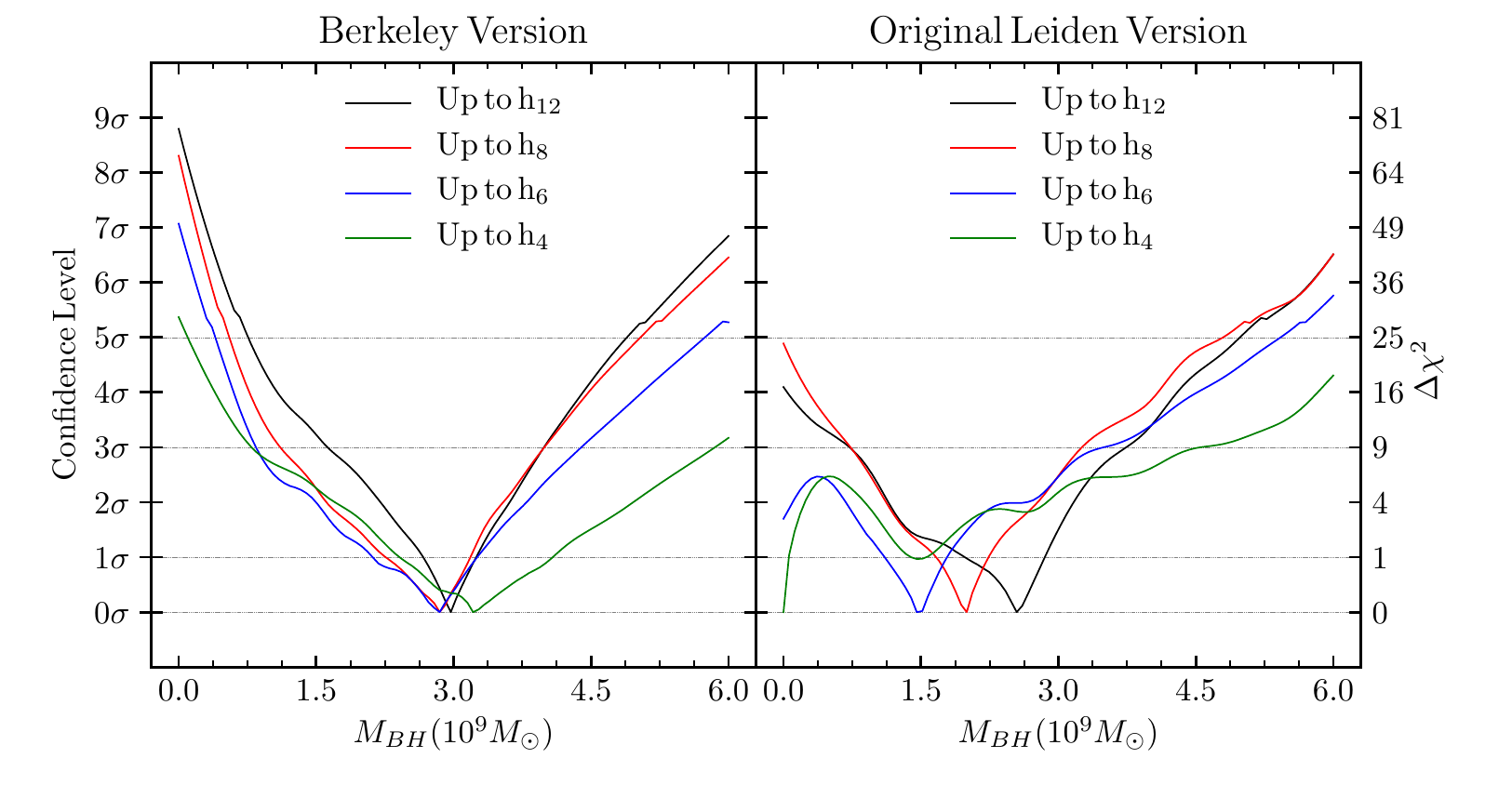}
  \caption{Illustration of the increasing constraints on \mbh\ provided by progressively higher Gauss-Hermite moments used to represent the LOSVDs. 
  Two versions of the orbit code with different settings are shown: the Berkeley version described in Sec.~4.1 (left), and the original triaxial Leiden version (right) with typical choice of $\psi$, including box orbits, and without orbit axisymmetrization.
  In each panel, we vary systematically the number of Gauss-Hermite moments used as constraints in the orbit model from 4 (green), 6 (blue), 8 (red), to 12 (black).
  Our production run corresponds to the black curve in the left panel. The green curve in the right panel uses a similar setting as in \citet{Ahnetal2018} for M59-UCD3 and prefers $\mbh=0$.
  The $\mbh=0$ minimum disappears as moments beyond $h_4$ are applied in the original code (right) but the best-fit \mbh\ is highly dependent on the truncation order. 
  In comparison, the main effect of additional moments in the Berkeley version (left) is to tighten the error bars on \mbh\ while leading the central value largely unchanged.
  Note we use a linear scale in the confidence level for the y-axis here for a clearer illustration of the locations of the minima, while Figure 8 uses a linear scale in $\chi^2$. 
  \label{figure:ChisqVsMbh_code_changes}}
\end{figure*}

The resulting first 12 moments predicted by the best-fit model for each of the four cases of increasing truncation orders are shown in Figure~\ref{figure:higher_moments}.
The corresponding LOSVDs for three representative GMOS bins in each case are shown in Figure~\ref{figure:LOSVDs}.
The marginalized $\chi^2$ versus \mbh\ for the four cases are shown in Figure~\ref{figure:ChisqVsMbh_code_changes} (left panel),
and the best-fit \mbh\ are listed in Table~\ref{tab:bestFit}
under ``Berkeley Version".
We note that while the best-fit \mbh\ changes by only $\sim 20$\%  (in the range of $2.6$ to $\SI{3.2e9}{M_\odot}$) as the truncation order is varied, the confidence level is improved significantly when more input moments are used, and the errors on \mbh\ are reduced by a factor of $\sim 2.3$ when we increase the truncation order from 4 to 12.

The detailed dependence of each of the 12 Gauss-Hermite moments on the truncation order can be clearly seen in Figure~\ref{figure:higher_moments}.  These moments are determined from the LOSVD of each bin (e.g., Fig.~\ref{figure:LOSVDs}) through the direct integration described by \citet{vanderMarelFranx1993}. Although only moments up to the truncation order are used for constraining the model LOSVDs, arbitrary higher moments can be computed.
The lowest 4 moments $V$, $\sigma$, $h_3$ and $h_4$ predicted by the best-fit models are mostly independent of the truncation order we tested.  This is not surprising since these 4 moments are fit during modelling in all cases.
The predicted moments beyond $h_4$, however, start to show varying degrees of deviations.
The case in which the series is truncated at $h_{12}$ (black points) corresponds to our production run.
It uses all 12 moments as constraints by design, so as expected, the best-fit model is well-behaved in all 12 panels.  
In comparison, when only 4 moments are fit by the orbit-based models (green points), the unconstrained 5th moment and beyond deviate strongly from the black points.  Similarly, when 6 (blue) or 8 (red) moments are used as constraints during the modelling, the 7th or 9th moment and beyond also show deviations from the black points.
Importantly, the deviation from 0 is not random; instead, the unconstrained moments are correlated spatially, being somewhat symmetric about $R=0"$ for even moments and antisymmetric about $R=0"$ for odd moments. The general trend that we observe in Figure~\ref{figure:higher_moments} is that the lower the truncation order is, the more their higher moments show unobserved and correlated features.

We illustrate the spurious features in the shapes of the LOSVDs resulting from the unconstrained higher Gauss-Hermite moments in Figure~\ref{figure:LOSVDs}. 
For all three representative GMOS bins shown, the model LOSVDs have the most pronounced irregular features when only 4 moments are used (green curve), and these features gradually go away as the truncation order is increased.

To date, a number of published dynamical \mbh\ measurements based on orbit modeling of stellar kinematics have used the method of Gauss-Hermite expansion to approximate the LOSVDs.
Most have used the first four moments as constraints in the orbit models,
e.g., \citet{Verolmeetal2002, vandenboschdezeeuw2010,vandenBoschetal2012, Walshetal2012, Walshetal2015, Walshetal2016, Walshetal2017, Ahnetal2018, Sethetal2014, Krajnovicetal2018, Thateretal2017,Thateretal2019}, while a few have used the first six moments, e.g., \citet{ Cappellarietal2002,Cappellarietal2009,Krajnovicetal2009}.
Our tests here are applied only to the triaxial Leiden code in the case of \ngc, so we cannot speak directly to the impact of higher Gauss-Hermite moments on \mbh\ in other work. However, we recommend that similar tests be performed in future work.

\subsection{The $M_{BH} = 0$ model: comparison to Ahn et al. (2018)}\label{BH_zero_solution}

Recently, \citet{Ahnetal2018} used the same triaxial orbit code as in this work to perform axisymmetric mass modeling of the ultracompact dwarf galaxy M59-UCD3.
They reported a puzzling global $\chi^2$ minimum at $\mbh = 0$, which was inconsistent with their best-fit $\mbh= 4.2^{+2.1}_{-1.7} \times 10^6 M_\odot$ from Jeans modeling and an orbit code that is intrinsically axisymmetric.  
Various tests were performed but none explained the discrepancy.
They speculated about a ``numerical artifact'' in the triaxial Leiden code and favored the non-zero \mbh\ from Jeans and axisymmetric orbit modeling.

As discussed in Sec.~4 and 7.1, we have made a number of changes to the original triaxial code and typical settings to arrive at the ``Berkeley version" results for \ngc\ presented in Sec.~5 and 6.  Even though our final outcome in Figure~\ref{figure:ChisqVsMbh} shows $\mbh = 0$ to be disfavored at the $8.7\sigma$ confidence level, we also encountered difficulties in constraining \mbh\ in the case of \ngc\ when we ran the original triaxial code using similar settings as those of \citet{Ahnetal2018}; that is, choosing $|\psi - 90^\circ| \ge 10^{-3}$, including box orbits from the default library, not axisymmetrizing the loop orbits, and using 4 Gauss-Hermite moments as kinematic constraints. Our resulting $\chi^2$ for this setting using the original code is represented by the green curve in the right panel of Figure~\ref{figure:ChisqVsMbh_code_changes}. 
The overall constraint on the \ngc\ \mbh\ is weak, with the lowest $\chi^2$ occurring at $\mbh=0$, and another local $\chi^2$ minimum at $\mbh\sim 1.5\times 10^{9} \msun$.  This is in stark contrast to the result from our version of the settings and code represented by the black curve
in the left panel of Figure~\ref{figure:ChisqVsMbh_code_changes}. 

In view of the importance of constraining higher Gauss-Hermite moments (Sec.~\ref{GH_discussion}), 
we have run further tests using the original code but increasing the number of input moments from 4 to 6, 8, and then 12.  The results are plotted in the right panel of
Figure~\ref{figure:ChisqVsMbh_code_changes}.
The $\chi^2$ minimum at $\mbh=0$ in the case of $h_4$ disappears
as the truncation order is increased,
but the location of the $\chi^2$ minimum depends sensitively on the number of moments, and the best-fit \mbh\ increases monotonically and shows no convergence even at order 12, as listed in Table~\ref{tab:bestFit}. In comparison, models from the ``Berkeley Version" in the left panel of
Figure~\ref{figure:ChisqVsMbh_code_changes} and in Table~\ref{tab:bestFit} have better-behaved $\chi^2$ contours. 

\subsection{Comparison to Jeans modeling}

In \citet{Eneetal2018}, we applied the method of Jeans Anisotropic Modeling (JAM; \citealt{cappellari2008}) to determine the mass parameters in \ngc.
JAM  is computationally cheap but is limited by the assumptions of axisymmetric potentials and cylindrically aligned velocity ellipsoids, and by the fact that its solutions could be unphysical. JAM has been shown to give consistent results as axisymmetric orbit models for regular fast rotators like \ngc\ (\citealt{Cappellarietal2010}). 

\begin{table}
\centering
\caption{Best-fit black hole mass and $1\sigma$ ($68\%$) confidence intervals for the eight cases shown in Figure~12. 
The 4-moment Leiden run results in a $\chi^2$
minimum at $\mbh=0$. For all other runs, the quoted confidence intervals here are properly marginalized through the routine described in Appendix~\ref{interpolation}.
\label{tab:bestFit}}
\begin{tabular}{ccc}
\# of constrained  & Berkeley Version & Original Leiden Version  \\
 moments & \mbh\ ($10^9 M_\odot$)& \mbh\ ($10^9 M_\odot$)   \\
\hline
4   &  $ 3.30 \pm 0.81 $ &   0 \\
6   &  $ 2.64 \pm 0.59 $ &  $ 1.51 \pm 0.49$             \\
8   &  $ 2.63 \pm 0.48 $ &  $ 1.93 \pm 0.41$          \\
12  &  $ 2.91 \pm 0.35 $ &  $ 2.22 \pm 0.55$            \\
\end{tabular}

\end{table}

Previous studies with JAM have typically assumed a globally constant $\beta_z = 1 - \langle v_z^2\rangle/\langle v_R^2 \rangle$, which
quantifies the flattening of the velocity ellipsoid along the minor axis.
In order to at least partially replicate orbit-type variation, we allowed two different values for $\beta_z$,
one for the Gaussian components with $\sigma_k < 1''$, and the other for the Gaussian components with $\sigma_k > 1''$.
The choice of $1''$ is motivated by the light profile of \ngc\, which starts to fall off more rapidly at $R \gtrsim 1''$ (see bottom panel of Fig.~\ref{fig:mge}). 

The JAM Modelling in \citet{Eneetal2018} used the distance 56.4 Mpc from the 2MASS redshift survey. After adjusting to 51.0 Mpc, the best-fit parameters from JAM are $\mbh = (2.98 \pm 0.23) \times 10^{9} \ M_{\odot}$ and $\ml = 2.28 \pm 0.04$ (with 1$\sigma$ errors). The \mbh\ value is within the 1$\sigma$ confidence interval of our best-fit value, while the \ml\ is higher than our best-fit value but is consistent with the 3$\sigma$ interval.

The best-fit circular velocity for the dark matter halo from JAM is $V_c = 364 \pm 45 \kms$ with $R_c = 6.5 \pm 2.5$ kpc.
Assuming that the uncertainties in these two parameters are highly correlated, this corresponds to $\mdm = (3.89\pm 0.96) \times 10^{11} M_\odot$.  This is roughly half our preferred value of $\mdm = (6.98\pm0.73) \times 10^{11} M_\odot$. At small radii, the enclosed dark matter in the JAM model is much larger than ours. The central density of JAM's halo is $(17.4\pm 4.3) \times 10^7 M_\odot / \si{kpc}^3$, while ours is half at $(9.9\pm 1.0) \times 10^7 M_\odot / \si{kpc}^3$. At $\SI{6.9}{kpc}$, the enclosed masses of the two halos are identical. 

The best-fit velocity anisotropy is
$\beta_z^G (\sigma_k^\prime < 1 '' ) = -0.58 \pm 0.62$ for the inner part and $\beta_{z}^G (\sigma_k^\prime > 1'') = 0.15 \pm 0.04$ for the outer part.  The anisotropy in the central region is comparable to what we find in Section~\ref{Orbital_structure}, but we find the the orbits to be much more radially biased in the outer region.  The black hole and stellar mass distribution of the JAM best-fit model and our best-fit model are very similar, suggesting that similar velocity anisotropies are required to fit the kinematics. Conversely, the enclosed mass due to the halo in our best-fit model is much larger than that of JAM beyond $\SI{6.9}{kpc}$, suggesting that our model must be more radially biased to similarly fit the kinematics, as we observe.

\subsection{Black Hole Scaling Relations}

The SMBH at the center of \ngc\ lies 0.32 dex above the mean \mbh-$\sigma$ scaling relation from \citet{McConnellMa2013}, which is within the 0.38 dex intrinsic scatter in that relation. For the other seven MASSIVE galaxies with stellar dynamical \mbh, four galaxies -- NGC~4472, NGC~4486, NGC~4649 and NGC~7619 -- are within 0.3 dex of the scaling relation, whereas the other three -- NGC~1600, NGC~3842, and NGC~4889 -- have \mbh\ that is overmassive by a factor of $\sim 3-6$ than predicted by their respective galaxy velocity dispersion.  These 8 galaxies exhibit similarly large scatter in the scaling relation between \mbh\ and the bulge stellar mass, where $M_*$ spans a factor of $\sim 3$ while \mbh\ spans a factor of $\sim 10$.

The stellar core radius of \ngc\ from our photometry is $r_b = 0.97''$ (0.24 kpc). This value is obtained by fitting a 2D core-S\'ersic profile, convolved with the PSF from \citet{Goullaudetal2018}. This fit was performed using Imfit \citep{Erwin2015}. The scaling relation between \mbh\ and $r_b$ for a sample of 21 massive cored ETGs is found to be $\log_{10} \mbh = 10.27 + 1.17 \log_{10}(r_b/{\rm kpc})$
with an intrinsic scatter of 0.29 dex \citep{Thomasetal2016}.  Our inferred \mbh\ for \ngc\ is only 0.077 dex below this relation.

\subsection{Gas Kinematics}

In \citet{Pandyaetal2017}, we observed the kinematics of warm ionized gas out to $\sim 8$ kpc within \ngc\, by tracing the $\SI{3727}{\angstrom}$ [O II] emission line using the spectra obtained with the Mitchell IFS. This warm gas was found to rotate with a ${\rm PA}_{\rm gas} \sim 312.5^\circ$, roughly perpendicular to the stellar rotation along ${\rm PA}_{\rm stars} \sim 35^\circ$. This extreme misalignment suggests that the warm gas originated from external accretion rather than in-situ stellar mass loss.  The gas was observed to have a rotation velocity of up to $\sim \SI{200}{\kms}$ and a comparable velocity dispersion, giving an rms velocity of $\sim 300 \kms$, similar to that of the stars studied in this paper. 

\section{Summary}

We have presented a black hole mass determination of the MASSIVE survey galaxy \ngc\ using high-spatial resolution stellar kinematic data from the GMOS IFS, wide-field kinematic data from the Mitchell IFS, and photometry from \emph{HST} WFC3.  Stellar kinematics are measured from the spectra to produce a truncated Gauss-Hermite parameterization of the LOSVDs. 
We determine the first eight moments of the LOSVDs from the high-$S/N$ GMOS spectra and the first six moments from the Mitchell spectra (Figs.~3 and 4).  The two sets of kinematic data together span about two orders of magnitude in radial extent, from $0.3"$ to $76"$ ($\sim 3$ effective radii) with a total of 173 spatial bins of varied size.

In the production run described in this paper, we perform axisymmetric Schwarzschild orbit modelling for more than 8000 mass models to determine the mass parameters in \ngc. 
For each mass model, we use a library of up to  800,000 stellar orbits to sample the phase space,
and then use a quadratic programming solver to find a superposition of orbits that minimizes the $\chi^2$ associated with the observed kinematics and also fit the observed photometry to within 1\%.
This procedure is done for all mass models to produce likelihood distributions for the mass parameters (Figs.~7 and 8).
The best-fit model for \ngc\ has a black hole mass $\mbh = (2.9\pm 0.4)\times 10^9 M_\odot$, a stellar mass-to-light ratio (in F110W band) $\ml = (2.09\pm 0.06) M_\odot / L_\odot$, and an enclosed dark matter mass $\mdm = (7.0 \pm 0.7)\times 10^{11} M_\odot$ at 15 kpc. The inclination is found to be nearly edge-on (Fig.~6).

We began the orbit modeling with the original triaxial Schwarzschild code of \citet{vandenBoschetal2008} but determined that numerous changes must be made to properly model axisymmetric systems with that code.  We found the gravitational potential not to be sufficiently axisymmetric when we adopted the typical setting of this code used in prior studies.
As a result, the orbit start space includes box and long-axis orbits that are forbidden in truly axisymmetric potentials. Additionally, many of the integrated orbits near the black hole or far into the halo do not exhibit axisymmetry as their precession timescale is much longer than the code's default integration time. We introduced an additional axisymmetrizing step to enforce this symmetry. 
We also addressed several other issues and improved the computational efficiencies in the code.
The changes leading to the Berkeley version of  the code is discussed in  Sec.~4; further details are described in Quenneville et al., in prep.

Another key finding of this paper is that care must be taken to properly handle the truncation of the Gauss-Hermite series used to describe the stellar LOSVDs. When the higher-order terms in the series are left unconstrained for \ngc, the resulting best-fit LOSVDs produced by the orbit models contain spurious features (Fig.~11), and the contributions from the unconstrained higher-order moments are not random but show spatial correlations (Fig.~10)
When the Berkeley version of the orbit code is used, we find that the confidence level on the \mbh\ determination for \ngc\ is significantly improved when at least 8 Gauss-Hermite moments are used as constraints: the 1-sigma confidence interval shrinks by a factor of $\sim 2$ relative to models with typical constraints on only $V$ through $h_4$ (left panel of Fig.~12).  By contrast, the $\chi^2$ landscape is not as well behaved when the original code is used with typical settings (right panel of Fig.~12).
Tests on each individual galaxy would have to be performed to assess whether earlier \mbh\ determinations are similarly impacted.

A number of the findings and code changed discussed in this paper are also
relevant when the orbit code is applied to a triaxial gravitational potential.
In particular, the problem of insufficient integration time for the subset of orbits with long precession timescales occurs in both 
axisymmetric and triaxial models.
We are currently investigating these issues with the aim to build equilibrium triaxial models for non-axisymmetric galaxies.

\acknowledgments
The MASSIVE survey is supported in part by NSF AST-1411945, NSF AST-1411642, NSF AST-1815417, NSF AST-1817100, HST GO-14210, HST GO-15265 and HST AR-14573. M.E.Q. acknowledges the support of the Natural Sciences and Engineering Research Council of Canada (NSERC), PGSD3-517040-2018. C.-P.M. acknowledges support from the Heising-Simons Foundation, the Miller Institute for Basic Research in Science, and the Aspen Center for Physics, which is supported by NSF grant PHY-1607611.  J.L.W. is supported in part by NSF grant AST-1814799.  This work used the Extreme Science and Engineering Discovery Environment (XSEDE) at the San Diego Supercomputing Center through allocation AST180041, which is supported by NSF grant ACI-1548562.  This work is based on observations obtained at the Gemini Observatory, processed using the Gemini IRAF package, which is operated by the Association of Universities for Research in Astronomy, Inc., under a cooperative agreement with the NSF on behalf of the Gemini partnership: the National Science Foundation (United States), National Research Council (Canada), CONICYT (Chile), Ministerio de Ciencia,
Tecnología e Innovación Productiva (Argentina), Ministério da Ciência, Tecnologia e Inovação (Brazil), and Korea Astronomy and Space Science Institute (Republic of Korea).

\appendix

\section{Multi-Gaussian Expansion parameters}\label{MGE_parameters}

We list in Table~\ref{tab:mge} the best-fit parameters of the 10 MGE components to our \emph{HST} WFC3 IR photometry of NGC~1453 shown in Figure \ref{fig:mge}.  The 10 Gaussians terms are assumed to have the same center and position angle of $28.5^\circ$.

We also performed an MGE fit to this photometry in \citet{Eneetal2018}. The two MGEs differ in that the fit presented here was found by using the 'mge\_fit\_sectors()' function rather than the 'mge\_fit\_sectors\_regularized()' function. We find that when the regularized fit is performed, the photometry is similarly well-fit. However, for the regularized fit we find a significant uptick in the model's surface brightness in the central $\sim 0.1$ arcsec, below the pixel scale of the photometry. To avoid this un-physical feature, we use the un-regularized fit here.

\begin{table}[h]
\centering
\caption{Best-fit MGE parameters to the \ngc\ \emph{HST} WFC3 IR photometry. Each Gaussian component is parametrized by a central surface density $I_k = L_k/2\pi \sigma_k^{\prime 2} q_k^\prime$ (calculated using an absolute solar AB magnitude $M_{\odot{\rm, F110W}}=4.54$), dispersion $\sigma_k^{\prime}$ (in arcseconds), and axis ratio $q_k^{\prime}$.}

\begin{tabular}{c|c|c}
$I_k \ \ [L_{\odot} / {\rm pc}^{2}]$ & $ \ \ \sigma_k^{\prime} \ \ [''] \ \ $ & $ \ \ q_k^{\prime} \ \ $ \\
\hline
 $6285.72$   & $ \ 0.118 \ $   &  $ \ 0.895 \ $ \\
 $11089.5$   &    $0.323$      &     $0.928$ \\ 
 $15865.7$   &    $0.715$      &     $0.863$ \\
 $9393.34$   &    $1.392$       &     $0.794$ \\  
 $5676.12$   &    $2.373$       &     $0.852$ \\
 $1824.78$   &    $3.846$       &     $0.791$ \\
 $1326.46$   &    $5.962$       &     $0.848$ \\  
 $561.023$   &    $10.501$       &     $0.786$ \\
 $280.091$   &    $20.747$       &     $0.823$ \\
 $80.423$    &    $47.289$       &     $0.896$ \\
\end{tabular}
\label{tab:mge}
\end{table}

\section{Interpolation and Marginalization} \label{interpolation}
We perform an interpolation with cubic radial basis functions (RBF) to promote our discrete sample of $\chi^2$ evaluations at each model point to a continuous function over the parameter-space. We use a variation on the implementation described by \citet{Knyshetal2016}. The RBF interpolation is described by
\[\chi^2(\vec{x}) = \sum_{i=1}^N \lambda_i (||T (\vec{x} - \vec{x}_i)||)^3 + \vec{b} \cdot \vec{x} + a,\]
where $\vec{x}$ describes a point in the parameter-space and $\lambda_i$, $\vec{b}$, and $a$ are uniquely defined from the criterion that the interpolation passes through all $N$ sample points. $T$ is initially the identity matrix.

A spatial rescaling is performed to improve the fit around the minimum of the landscape. This is done by evaluating the interpolating function at $10000$ points drawn from a uniform distribution over the parameter-space. The covariance matrix of the $500$ points with lowest predicted $\chi^2$ is computed, then the eigenvalues $\alpha_i$ and eigenvectors $\vec{m}_i$ of that matrix are computed. Finally, $T$ is constructed with $\vec{T}_i = \vec{m}_i / \sqrt{\alpha_i}$. Given this new $T$, $\lambda_i$, $\vec{b}$, and $a$ are recomputed so that the interpolation once again passes through all $N$ sample points.

To extract best-fit values and confidence intervals for each parameter we perform a straightforward marginalization. With marginalization we wish to reduce the interpolated $\chi^2(\vec{\theta},\vec{\psi})$ to $\chi^2(\vec{\theta})$, where $\vec{\psi}$ are the parameters we wish to eliminate and $\vec{\theta}$ are those which remain. The likelihood is related to the $\chi^2$ by $L = e^{-\chi^2 / 2}$ and likelihoods are marginalized in the same sense as probabilities. Therefore
\[L(\vec{\theta}) = \int d^N \psi L(\vec{\theta},\vec{\psi})\]
and thus
\[\chi^2(\vec{\theta}) = -2 \ln \int d^N \psi e^{-\chi^2(\vec{\theta},\vec{\psi}) / 2}\],
where $N$ is the number of parameters in $\psi$. 

To obtain predictions for the best-fit and confidence interval for a parameter, we first construct the 1D likelihood function for that parameter:
\[L(\theta) = \int d^N \psi e^{-\chi^2(\theta,\vec{\psi}) / 2}.\]
For the best-fit, we determine the value where the cumulative-likelihood function is one-half:
\[\frac{\int_{-\infty}^\theta L(\theta') d\theta'}{\int_{-\infty}^\infty L(\theta') d\theta'} = \frac{1}{2}.\]
For the confidence intervals, we find the values where the cumulative-likelihood function reaches reaches the appropriate percentiles:
\[\frac{\int_{-\infty}^{\theta_\pm} L(\theta') d\theta'}{\int_{-\infty}^\infty L(\theta') d\theta'} = \frac{1 \pm \text{erf}(k / \sqrt{2})}{2},\]
where $\theta_+$ and $\theta_-$ yield the upper and lower bounds to the cumulative-likelihood and $k$ sets the confidence level ($k = 1$ corresponds to the $68\%$ level, $k = 2$ for $95\%$, and so on).
We compute these integrals with the VEGAS Monte Carlo integrator implemented in the Python package 'vegas'.

\end{document}